\documentclass[prb,reprint,superscriptaddress]{revtex4-1}

\usepackage{amsmath}
\usepackage{amssymb}

\usepackage{physics}	

\usepackage{comment}
\usepackage{subfig}

\usepackage{graphicx}

\usepackage{hyperref}
\hypersetup{
 bookmarks=true,		
 unicode=false,			
 pdftoolbar=true,		
 pdfmenubar=true,		
 pdffitwindow=false,		
 pdfstartview={FitH},		
 pdftitle={Correlation analysis of electron-deficient bonds},    
 pdfauthor={Jan Brandejs} {Libor Veis} {Szil{\'a}rd Szalay} {Gergely Barcza} {Ji{\u r}{\'i} Pittner} {\"Ors Legeza},	
 pdfsubject={chemical bonds are identified with strong orbital correlation},	
 pdfcreator={pdflatex},		
 pdfproducer={vim},		
 pdfkeywords={quantum chemistry} {chemical bond} {electron-deficit bonds} {quantum correlations} {entanglement} {multiorbital correlations} {diborane} {C-Be-C complex}, 
 pdfnewwindow=true,		
 colorlinks=true,		
 linktoc=page,			
 linkcolor=blue,		
 citecolor=blue,		
 filecolor=blue,		
 urlcolor=blue			
}
 \usepackage[usenames,dvipsnames]{xcolor}
 \usepackage[normalem]{ulem}
% ...

\newcommand{\updownarrows}{\uparrow\mathrel{\mspace{-1mu}}\downarrow}

\newcommand{\rest}{\text{rest}}

\newcommand{\atom}[2]{{\mathrm{#1}^{#2}}}

\newcommand{\orbital}[2]{{\mathrm{#2}}}

\newcommand{\orbitalh}[2]{{\mathrm{sp}^3_{#2}}}

\newcommand{\orb}[4]{\atom{#1}{#2}\orbital{#3}{#4}}

\newcommand{\orbh}[4]{\atom{#1}{#2}\orbitalh{#3}{#4}}

\newcommand*{\citen}[1]{%
  \begingroup
    \romannumeral-`\x 
    \setcitestyle{numbers}%
    \cite{#1}%
  \endgroup   
}

\begin{document}

\title{Quantum information-based analysis of electron-deficient bonds}
\author{Jan Brandejs}
\email{jan.brandejs@jh-inst.cas.cz}
\affiliation{J. Heyrovsk\'{y} Institute of Physical Chemistry, Academy of Sciences of the Czech \mbox{Republic, v.v.i.}, Dolej\v{s}kova 3, 18223 Prague 8, Czech Republic}
\affiliation{Faculty of Mathematics and Physics, Charles University, Prague, Czech Republic}
\author{Libor Veis}
\email{libor.veis@jh-inst.cas.cz}
\affiliation{J. Heyrovsk\'{y} Institute of Physical Chemistry, Academy of Sciences of the Czech \mbox{Republic, v.v.i.}, Dolej\v{s}kova 3, 18223 Prague 8, Czech Republic}
\author{Szil\'ard Szalay}
\email{szalay.szilard@wigner.mta.hu}
\affiliation{Strongly Correlated Systems ``Lend{\"u}let'' Research Group,
Institute for Solid State Physics and Optics,
MTA Wigner Research Centre for Physics,
H-1121 Budapest, Konkoly-Thege Mikl{\'o}s {\'u}t 29-33, Hungary}
\author{Gergely Barcza}
\email{barcza.gergely@wigner.mta.hu}
\affiliation{J. Heyrovsk\'{y} Institute of Physical Chemistry, Academy of Sciences of the Czech \mbox{Republic, v.v.i.}, Dolej\v{s}kova 3, 18223 Prague 8, Czech Republic}
\affiliation{Strongly Correlated Systems ``Lend{\"u}let'' Research Group,
Institute for Solid State Physics and Optics,
MTA Wigner Research Centre for Physics,
H-1121 Budapest, Konkoly-Thege Mikl{\'o}s {\'u}t 29-33, Hungary}
\author{Ji{\u r}{\'i} Pittner}
\email{jiri.pittner@jh-inst.cas.cz}
\affiliation{J. Heyrovsk\'{y} Institute of Physical Chemistry, Academy of Sciences of the Czech \mbox{Republic, v.v.i.}, Dolej\v{s}kova 3, 18223 Prague 8, Czech Republic}
\author{\"Ors Legeza}
\email{legeza.ors@wigner.mta.hu}
\affiliation{Strongly Correlated Systems ``Lend{\"u}let'' Research Group,
Institute for Solid State Physics and Optics,
MTA Wigner Research Centre for Physics,
H-1121 Budapest, Konkoly-Thege Mikl{\'o}s {\'u}t 29-33, Hungary}

\begin{abstract}
Recently, the correlation theory of the chemical bond was developed, which applies concepts of quantum information theory for the characterization of chemical bonds, based on the multiorbital correlations within the molecule. Here
for the first time, we extend the use of this mathematical toolbox for the description of electron-deficient bonds.
We start by verifying the theory on the textbook example of a molecule with three-center two-electron bonds, namely the diborane(6).
We then show that the correlation theory of the chemical bond is able to properly describe bonding situation in more exotic molecules which have been synthetized and characterized only recently, in particular the diborane molecule with four hydrogen atoms [diborane(4)] and neutral zerovalent s-block beryllium complex, whose surprising stability was attributed to a strong three-center two-electron $\pi$ bond stretching across the C-Be-C core.
Our approach is of a high importance especially in the light of a constant chase after novel compounds with extraordinary properties where the bonding is expected to be unusual.
\end{abstract}

\maketitle


\section*{Introduction}

Recent years have witnessed remarkable interest in application of tools of quantum information theory in chemistry \cite{Legeza-2003b, Legeza-2004b, Huang2005, rissler_2006, Pipek-2009, Barcza-2011, McKemmish-2011, boguslawski_2012, boguslawski_2012b, Boguslawski-2013, Kurashige-2013, Fertitta-2014, Duperrouzel-2014, murg_2014, Knecht-2014, Boguslawski-2015, Szalay-2015a, Boguslawski-2015, Freitag-2015, Zhao-2015, Szilvasi-2015, Molina2015, krumnow_2016, stein_2016, stein_2017, kovyrshin_2017, Szalay-2017, Stemmle2018}.
As a prominent example, the performance of state-of-the-art tensor product methods for electronic structure calculations \cite{Kurashige-2009, Murg-2010a, Nakatani-2013, Szalay-2015a, Chan-2015, Reiher-MPO, Wouters-2014e, Gunst-2018} heavily relies on proper manipulation of entanglement \cite{Legeza-2003b, rissler_2006, Barcza-2011, Fertitta-2014, murg_2014, Szalay-2015a, krumnow_2016}. These include
density matrix renormalization group (DMRG) method \cite{White-1992b, white_1993}, which variationally optimizes wave functions in the form of matrix product states (MPS).\cite{schollwock_2011} 

Other important examples represent characterization of electron correlation into its static (strong) and dynamic contributions \cite{boguslawski_2012b}, automatic (black-box) selection of the active spaces \cite{Legeza-2003b, Barcza-2011, Szalay-2015a, stein_2016, stein_2017, Faulstich2018}, or the self-adaptive tensor network states with multi-site correlators \cite{kovyrshin_2017},
all of which harness single- and two-orbital entanglement entropies.
Last but not least, correlation measures based on the single- and two-orbital entanglement entropies have also been employed for the purposes of bond ana\-ly\-sis \cite{Boguslawski-2013, Szilvasi-2015}.

In the preceding work \cite{Szalay-2017}, we have presented the very general \emph{correlation theory of the chemical bond} based on multiorbital correlation measures which goes beyond the scope of two-orbital picture. It is able to properly describe multiorbital bonds, and we have demonstrated its performance on a representative set of organic molecules (aliphatic as well as aromatic).

In the present article, we apply this theory to systems with electron-deficient bonds, i.e., to compounds which have too few valence electrons for the connections between atoms to be described as covalent bonds, and which have always fascinated chemists. First we apply the theory to the notoriously known textbook example of the diborane(6)\footnote{The number in parentheses denotes the number of hydrogen atoms.}
molecule (B$_2$H$_6$) with two-electron three-center bridge bonds and then also to recently characterized diborane(4) \cite{Chou2015} (B$_2$H$_4$) and zero-valent complexes of beryllium \cite{Arrowsmith-2016, Brabec2018}. The neutral form of the latter compound exhibits surprising stability, which was attributed to a strong three-center two-electron $\pi$ bond stretching across the C-Be-C core \cite{Arrowsmith-2016}. Unlike in the previous study \cite{Szalay-2017}, here we work in the bigger detail in a sense that we also employ eigenstates of multiorbital reduced density matrices, which give us additional insights into the character of bonding.

The article is organized as follows: in Sec. II, we briefly present the studied systems, in Sec. III we review the main concepts of the theory of multiorbital correlations, Sec. IV presents the computational details and Sec.~V the results of our calculations which are followed by their discussion, the final Section closes with conclusions.

\section*{Studied systems}

\subsection*{Diboranes}
\begin{figure}[!ht]
  \subfloat[diborane(6)\label{d6}]{%
    \includegraphics[width=4cm]{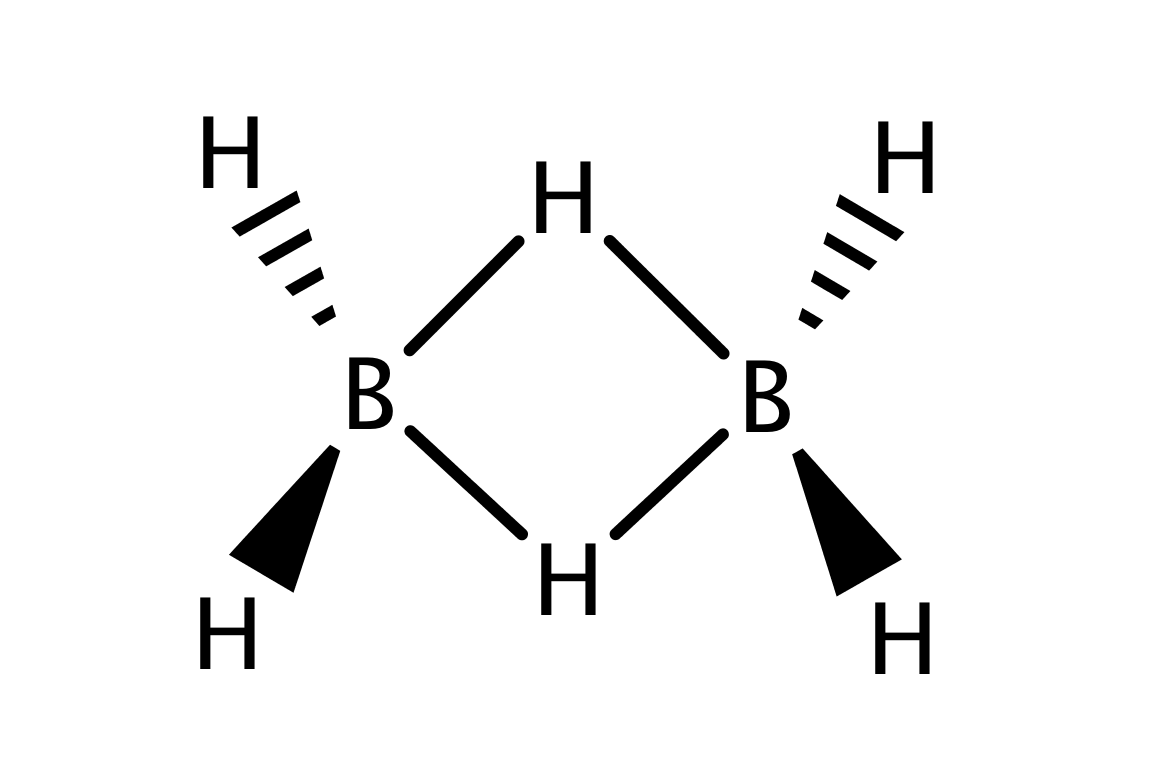}
  }
  \hfill
  \subfloat[diborane(4)\label{d4}]{%
    \includegraphics[width=4cm]{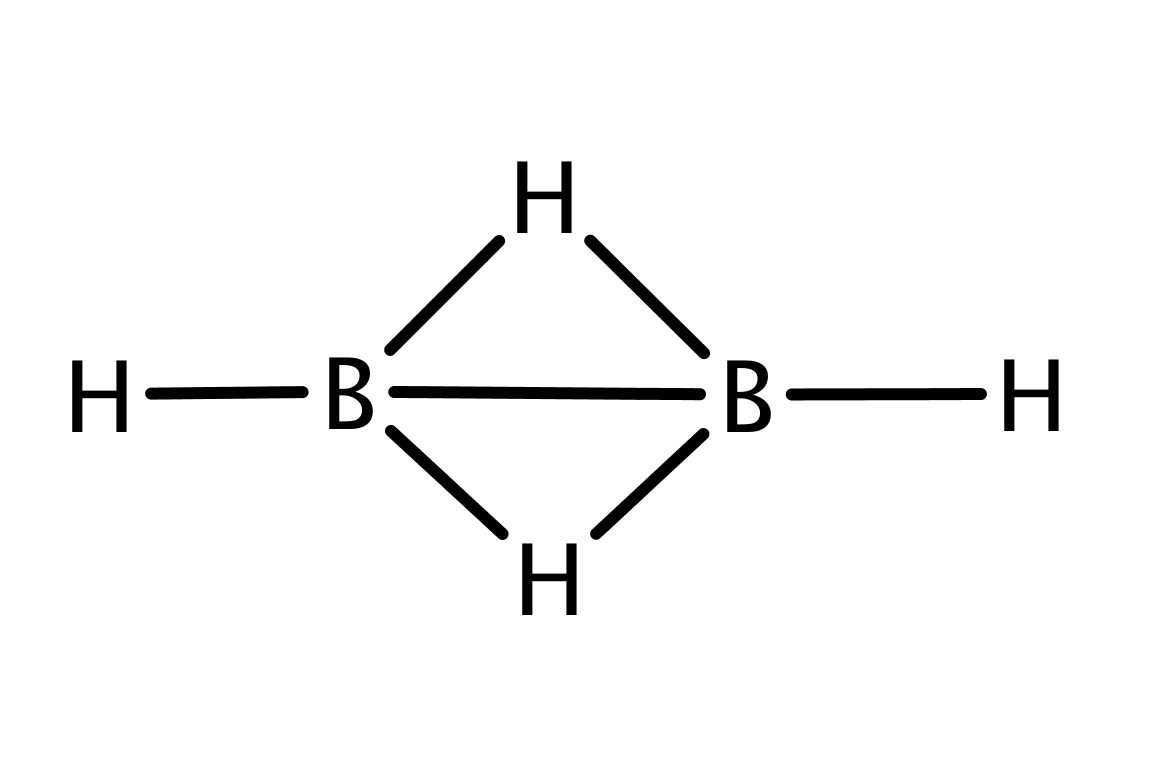}
  }
  \caption{Structures of diborane(6) and diborane(4).}
\end{figure}

In its ground state, diborane(6) (Figure \ref{d6}) adopts its most stable conformation with two 
bridging B-H-B bonds, and four terminal B-H bonds. Its structure was first 
correctly measured in 1943 from infrared spectra of gaseous samples 
by an undergraduate student, Longuet-Higgins.\cite{LonguetHiggins1946,LonguetHiggins1943}
Subsequent measurements with electron diffraction confirmed his conclusions\cite{Eberhardt1954}, and X-ray diffraction detected 
further systems with bridging hydrogen bond.\cite{Lammertsma1996}
The B-H-B bridging was considered an atypical electron-deficient covalent chemical bond.\cite{Lipscomb1973}
Diborane(6) is a prominent example of a molecule with three-center two-electron bonds.\cite{Neeve2016}  
As it is a well studied system, we use its B-H-B linkage as a reference to compare with 
bond strengths and properties of more complex systems featuring three-center two-electron bonds.

According to quantum-chemical calculations \cite{Vincent1981,Mohr1986,Curtiss1989,Curtiss1989b,Demachy1994,Alkorta2011}, different species of diborane 
with less than six hydrogens should exist, also featuring the bridging B-H-B bonds. 
However, all candidates are short-lived reaction intermediates, difficult to prepare and to identify.
Hence, no neutral species has been identified experimentally until 2015, when Chou irradiated 
diborane(6) dispersed in neon at $3\;\text{K}$ with far-ultraviolet light, detecting diborane(4), B$_{2}$H$_{4}$ (Figure \ref{d4}).\cite{Chou2015} 
This new species with two terminal hydrogen atoms possesses two bridging hydrogen atoms,
 and so it became the simplest neutral boron hydride identified with such a structural feature.

\subsection*{Beryllium complexes}
\begin{figure}
  \begin{center}
    \includegraphics[width=8cm]{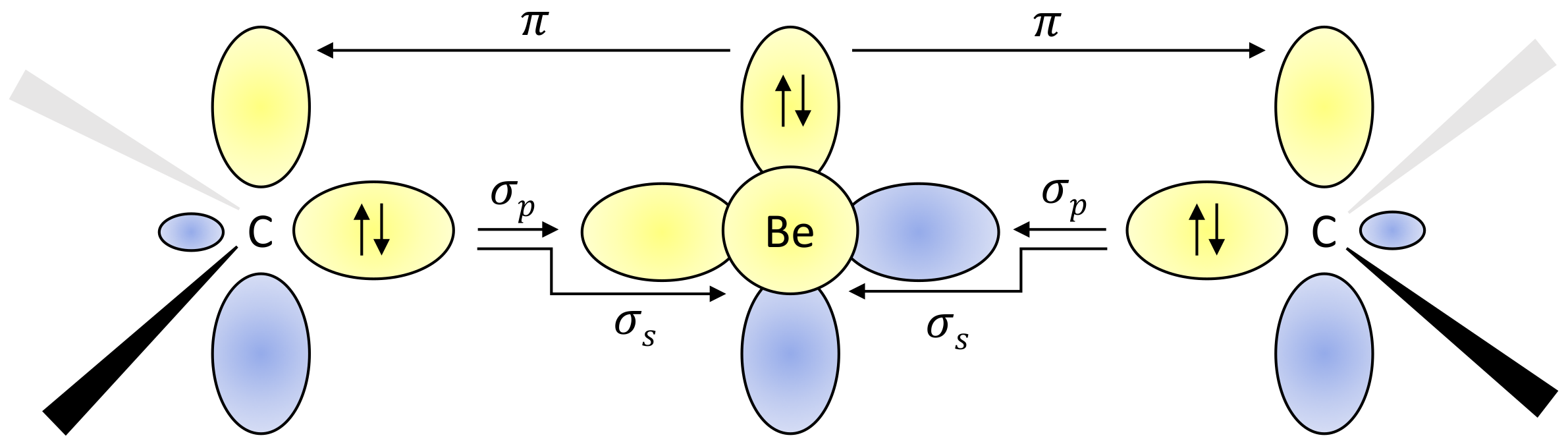}
    \caption{Schematic representation of the C-Be-C bonding mechanisms \cite{Arrowsmith-2016}.}
    \label{bonding_scheme}
  \end{center}
\end{figure}

Complexes of metal atoms of the s-block of the periodic table are often found 
in their zero oxidation state due to their exceptional electron donation. For 
their interesting reactivities, these became frequent synthetic targets, competing 
with traditional transition metal complexes.\cite{Power2010,Power2011,Giffin2011} 
We follow the recent experimental work of Arrowsmith, who isolated, for the first time, 
neutral compounds with zero-valent s-block metal, beryllium. \cite{Arrowsmith-2016}
These brightly coloured molecules have very short Be-C bonds and beryllium in 
linear coordination geometries.\cite{Niemeyer1997,Naglav2015,Arnold2015,Lerner2003} 
This indicates strong multiple Be–C bonding. 
According to the theoretical and spectroscopic results, the molecules adopt 
a closed-shell singlet configuration with a Be(0) metal centre. \cite{Arrowsmith-2016}
The complexes are surprisingly stable, and this was ascribed to an unusually 
strong three-center two-electron $\pi$ bond stretching across the C–Be–C unit.
Two bonding mechanisms depicted in Figure \ref{bonding_scheme} are taking place, namely $\sigma$
donation from the carbon doubly occupied sp$^2$ hybrid orbital to empty s and p$_\text{x}$
orbitals on the central Be atom and $\pi$ back donation from the beryllium p$_\text{z}$ orbital
to p$_\text{z}$ orbitals located on C atoms.

We studied two of the proposed systems. First the Be(CAC)$_{2}$ complex (Figure \ref{str}), where CAC 
corresponds to cyclic amino carbene donors, which stabilize the compound 
due to their $\pi$-acidity. \cite{Mondal2013,Li2013} We performed a multireference calculation, 
in order to verify the proposed singlet configuration with a Be(0) metal centre 
and to provide a deeper insight into the bonding scheme.
Next we studied dication [Be(CAC)$_{2}$]$^{2+}$ (Figure \ref{str2+}), in which the removal of two electrons 
disrupts the bridging C–Be–C bond. This allowed us to compare with the former system 
and to determine the stabilization effect of the bridging bond.

\section*{Methodology}

Recently, the correlation theory of the chemical bond \cite{Szalay-2017} was developed, characterizing bonds based on the correlations among orbitals localized on individual atoms. Simply put, if we think of a simple covalent bond and localize the bonding and antibonding molecular orbitals into their atomic contributions, these localized orbitals will be highly correlated. Therefore, standard two-orbital bonds can be characterized by pairs of strongly correlated localized orbitals, and the strength of the correlation characterizes the strength of the bond from the quantum information theoretical point of view.

The correlation theory of the chemical bond can also be used for the characterization of bonds more involved than the covalent bonds.
The concept in general is to find the finest possible correlation based clustering of the localized orbitals into clusters, so that the clusters are weakly correlated with each other, and the orbitals inside the clusters are strongly correlated.\cite{Szalay-2017} These clusters then form independent bonds of a Lewis structure of a given molecule 
and the strength of the correlation with respect to this clustering refers to the validity of such a representation. The weaker the correlation is, the better the Lewis structure represents bonding.

In order to review the correlation measures\cite{Szalay-2015b}, which will be used in our ana\-ly\-sis,
let us denote the set of (the labels of) localized orbitals with $M$.
We aim at investigating the correlations in an $L\subseteq M$ set of orbitals (cluster).
The state of the full electronic system of the cluster $L$
is given by the density operator $\varrho_L$,
while
the reduced state of a (sub)cluster $X\subseteq L$
is given by the reduced density operator $\varrho_X$
in general.\cite{Ohya-1993,Araki-2003b,Wilde-2013}
If the cluster of orbitals $L$ can be given by a state vector $\ket{\psi_L}$
(for example, when a given eigenstate of the whole molecule is considered),
then its density operator is of rank one, $\varrho_L=\ket{\psi_L}\bra{\psi_L}$, called a pure state.
Its reduced density operator is usually mixed (not of rank one),
which is the manifestation of entanglement\cite{Horodecki-2009} 
between (sub)cluster $X$ and the rest of the cluster $L\setminus X$.
In general, a density operator $\varrho_X$ can be decomposed in infinitely many ways
into state vectors $\ket{\psi_i}$ with mixing weights $p_i\geq0$ as
$\varrho_X=\sum p_i \ket{\psi_i}\bra{\psi_i}$.
The spectral decomposition (where the weights are the $\lambda_i$ eigenvalues,
and the $\ket{\psi_i}$-s are eigenvectors, being \emph{orthogonal})
is a special one, in the sense that its weights are the least mixed.\cite{Schrodinger-1936,Hughston-1993} 
Each eigenvector $\ket{\psi_i}$ can be expanded in the occupation number basis,
the square of the absolute value of the coefficients are the weights of the given occupations
in that given eigenvector $\ket{\psi_i}$ \mbox{of weight $\lambda_i$}.

On the first level,
the correlation is defined with respect to a partition\cite{Davey-2002} of the $L$ set of the orbitals,\cite{Szalay-2012,Szalay-2015b,Szalay-2017,Szalay2018cc}
denoted with $\xi =\{X_1,X_2,\dots,X_{\abs{\xi}}\} \equiv X_1|X_2|\dots|X_{\abs{\xi}}$,
where the clusters $X\in\xi$, called parts, are disjoint subsets of the cluster $L$,
and $\cup_{X\in\xi} X = L$.
The measure of correlation among the parts $X\in\xi$ is the
$\xi$-correlation,\cite{,Szalay-2015b,Szalay-2017}
\begin{equation}
\label{eqr:xiCorr}
	C(\xi) := \sum_{X\in\xi} S(X) - S(L).
\end{equation}
Here $S(X)=-\tr(\varrho_X\log_4\varrho_X)$ is the von Neumann entropy.\cite{Ohya-1993,Wilde-2013}
(Note that we use the logarithm to the base $4$, 
which is the dimension of the Hilbert space of an orbital. The resulting numerical values are then the same as of the original measures with natural logarithm\cite{Szalay-2017} given in the units of $\ln4$. Note that $S(X)\leq |X|$, where $|X|$ is the number of orbitals in cluster $X$.)
As a special case, the correlation of two single orbitals,
\begin{equation}
	\label{eqr:twoorbitMutInf}
	C(i|j) = S(i) + S(j) - S(i,j) = I(i|j),
\end{equation}
is the well-known (two-orbital) mutual information,\cite{Ohya-1993,Wilde-2013,Adesso-2016}
which has already been considered in chemistry.%
\cite{Legeza-2006a,Rissler-2006,Barcza-2011,Boguslawski-2013,Kurashige-2013,Mottet-2014,Fertitta-2014,Duperrouzel-2014,Boguslawski-2015,Freitag-2015,Szilvasi-2015,Murg-2015,Barcza-2015,Zhao-2015,Szalay-2017}
(For convenience, we omit the curly brackets $\{\;\}$ in the cases 
when this does not cause confusion.)
For a general partition $\xi$, we have the bound\cite{Szalay-2017} 
\begin{subequations}
	\begin{equation}
		C(X|Y)\leq 2 (|L| -  \max_{X\in\xi} |X|),
	\end{equation}
	which for a bipartition $\xi=X|Y$ reduces to 
	\begin{equation}
		C(X|Y)\leq 2 \min\{|X|,|Y|\}.
	\end{equation}
\end{subequations}

Note that $C(\xi)$ 
is zero for the trivial split $\xi=\top=\{L\}$,
and it takes its maximum, $C(\bot)$, for the finest split $\xi = \bot = \{ \{i\} \;:\; i\in L \}$.
The latter quantity
is also called total correlation,\cite{Lindblad-1973,Horodecki-1994,Legeza-2004b,Legeza-2006b,Herbut-2004}
\begin{equation}
	\label{eqr:totalCorr}
	C_\text{tot}(L) := C(\bot)=\sum_{i\in L}S(i)-S(L).
\end{equation}
(Note that if cluster $L$ is described by a pure state then $S(L)=0$,
and the correlation is entirely quantum entanglement.\cite{Horodecki-2009,Modi-2010,Szalay-2015b}
Moreover, the correlation in a pure state with respect to a bipartition $\xi=X|(L\setminus X)$  
is just two times the usual entanglement entropy\cite{Bennett-1996,Nielsen-2000,Wilde-2013}
\begin{equation}
        \label{eqr:entEnt}
	C(X|(L\setminus X)) = 2S(X) = 2S(L\setminus X),
\end{equation}
because of the Schmidt decomposition of pure states.\cite{Schmidt-1907,Nielsen-2000,Wilde-2013})

On the second level,
the correlations can be defined
in an overall sense, that is, without respect to a given partition.\cite{Szalay-2017,Szalay-2015b,Szalay2018cc}
The $k$-partitionability correlation and
the $k$-producibility correlation, are\cite{Szalay-2017}
\begin{subequations}
\label{eqr:kppCorr}
\begin{align}
	C_\text{$k$-part}(L) &:= \min_{\xi:\; \abs{\xi}\geq k}C(\xi),\\
	C_\text{$k$-prod}(L) &:= \min_{\xi:\; \forall X\in\xi,\: \abs{X}\leq k }C(\xi),
\end{align}
\end{subequations}
for $1\leq k\leq\abs{L}$.
These characterise the strength of two different (one-parameter-) notions of multiorbital correlations;
those which cannot be restricted inside at least $k$ parts,
and
those which cannot be restricted inside parts of size at most $k$, respectively.\cite{Szalay-2017}

For the cluster $L$,
as special cases,
$C_\text{$\abs{L}$-part}=C_\text{$1$-prod}=C(\bot)$ grabs all the correlations,
it is zero if and only if there is no correlation at all in the cluster $L$.
On the other hand,
$C_\text{$2$-part}=C_\text{$(\abs{L}-1)$-prod}$ is sensitive only for the strongest correlations,
it is nonzero if and only if the cluster $L$ is globally correlated.
Note also that $C_\text{$1$-part}=C_\text{$\abs{L}$-prod}=C(\top) = 0$, by definition.
Beyond these,
there are no such coincidences among the partitionability and producibility correlations
for other values of $k$,
however, the relation $C_\text{$k$-part}\geq C_\text{$(\abs{L}-k+1)$-prod}$ holds.\cite{Szalay-2017}
Also, the following (non stricht) bounds hold \cite{Szalay-2017}
\begin{subequations}
\label{eqr:bounds}
\begin{align}
	C_\text{$k$-part} &\leq 2(k-1), \\
	C_\text{$k$-prod} &\leq 2(\abs{L}-k).
\end{align}
\end{subequations}

\section*{Computational details}

In case of diborane molecules, 
the ground state geometries were optimized with the B3LYP/cc-pVDZ method. For the multiorbital correlation studies, the Pipek-Mezey \cite{Pipek-1989} localized HF/STO-3G molecular orbitals (MOs) were employed \cite{Szilvasi-2015, Szalay-2017} and they were manually hybridized (rotated) to better reflect the chemical environment. 
The quantum chemical (QC-) DMRG method was applied to study the multiorbital correlations in the full orbital following the procedure
outlined in Ref.~\citen{Szalay-2017}.

The ground state geometries of both forms of the beryllium complex, namely Be(CAC)$_2$ and [Be(CAC)$_2$]$^{2+}$ were taken from Ref.~\citen{Arrowsmith-2016} and they correspond to the BP86/def2-TZVPP level of theory. Due to the size of the problem, multiorbital correlation studies by means of QC-DMRG are not feasible in the full orbital space. We have rather chosen a different strategy. Since we were interested only in the bonding of the C-Be-C atomic core, we have selected the complete active space (CAS) of relevant orbitals participating or influencing these bonds. In particular, 2p$_{\text{x}}$ orbitals on both C and Be atoms and 2s orbital on Be, all of them contributing to the $\sigma$ bonds and 2p$_{\text{z}}$ orbitals on both C and neighbouring N atoms and Be, forming or directly influencing the $\pi$ bonds. The CAS orbitals were optimized by means of the CASSCF(10,9)/cc-pVDZ method in case of the neutral complex and CASSCF(8,9)/cc-pVDZ method in case of the dication and again localized using the Pipek-Mezey \cite{Pipek-1989} procedure. They were not hybridized in order to directly compare with the previous work \cite{Arrowsmith-2016} making conclusions about atomic-like orbitals.

All the quantum chemistry calculations except the QC-DMRG ones were performed with the MOLPRO package \cite{MOLPRO}.
The QC-DMRG calculations were carried out using the Budapest QC-DMRG code \cite{budapest_qcdmrg}. Molecular orbitals were visualized with Charmol \cite{charmol}.

\section*{Results}

The results on diborane(6) are summarized in Figure \ref{fig:B2H6} and Table \ref{table:B2H6}, whereas results on diborane(4) are presented in Figure \ref{fig:B2H4} and Table \ref{table:B2H4}. All the Figures depict mutual information of pairs of localized orbitals, defined in \ref{eqr:twoorbitMutInf}, while the Tables contain numeric values of measures of the relevant kinds of multiorbital correlations,  which are discussed below.

In a similar fashion, the results on beryllium complexes are presented in Figures \ref{fig:CBeCv2}, \ref{fig:dication}, and \ref{fig:cbecrotatedsmall} and Table \ref{table:cbec01}.

In the Figures, 
individual localized orbitals are 
represented as black dots and dashed blue lines encircles orbitals belonging to one atom.
The mutual information is plotted as grayscaled edges between the orbitals. Black lines correspond to the strongest
correlations, while light gray lines connect the weakly correlated orbitals.
Based on the mutual information structure, the orbitals are grouped into strongly correlated clusters, which in our examples correspond to either core orbitals or chemical bonds.
The clusters are encircled by red borders.

\section*{Discussion}

\subsection*{Diborane(6)}

\begin{figure}[!ht]
\centering
\includegraphics{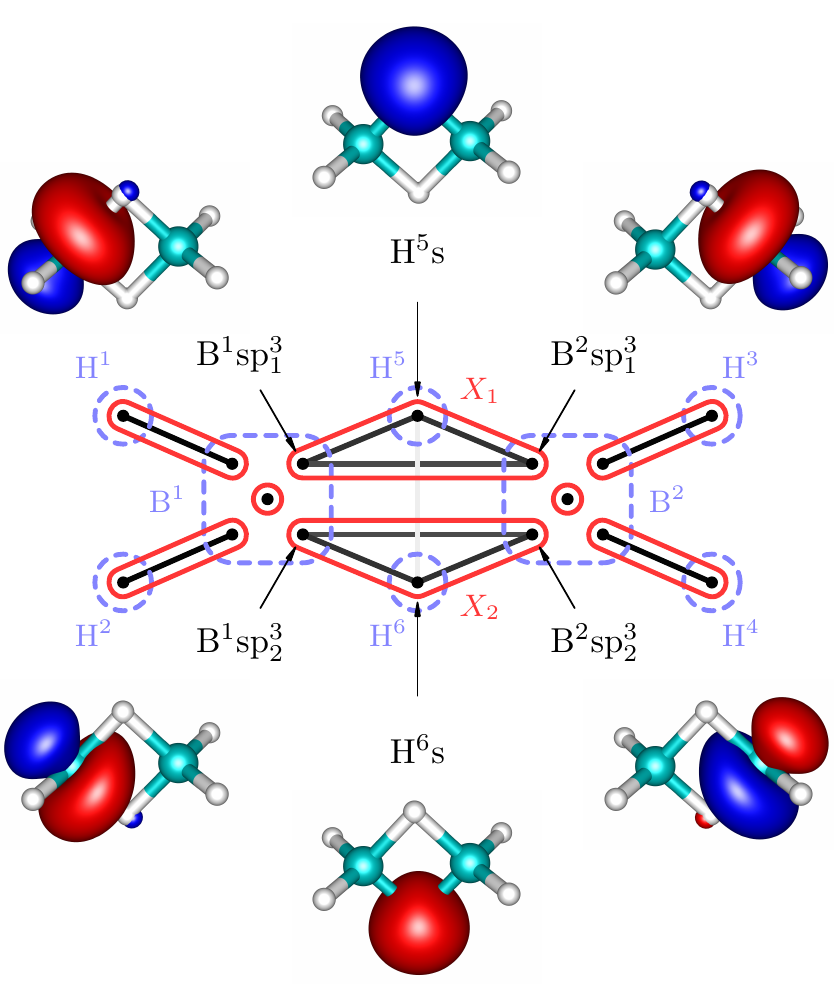}
\caption{Schematic view of diborane(6) with mutual information: 
	each dot represents a localized orbital, 
	dashed blue line encircles individual atoms,
	edges correspond to mutual information (plot shaded by a 
	logarithmic scale depending on strength) and 
red circles show how the orbitals group into clusters, i.e. independent bonds. Sorted values of the two-obital mutual information are plotted in Appendix \ref{appendix}.}
\label{fig:B2H6}
\end{figure}

\begin{table}[!ht]
\centering
\begin{tabular}{l c c}
 \hline
 \hline
 correlation & abs. value & rel. value \\ [0.5ex]
 \hline
 $C(X_1 \;|\; \rest)$ & 0.515 & 8.6\% \\
 $C(X_1 \cup X_2 \;|\; \rest)$ & 0.412 & 3.4\% \\
 $C(\orbh{B}{1}{2}{1}, \orb{H}{5}{1}{s} \;|\; \rest)$ & 1.852 &  46\% \\
 $C(\orb{H}{5}{1}{s},\orbh{B}{2}{2}{1} \;|\; \rest)$ & 1.852 & 46\% \\
 $C(\orbh{B}{1}{2}{2},\orbh{B}{2}{2}{1} \;|\; \rest)$ & 2.114 & 53\% \\
 $C(\orbh{B}{1}{2}{1} \;|\; \orb{H}{5}{1}{s})$ & 0.894 & 45\% \\
 $C(\orb{H}{5}{1}{s} \;|\; \orbh{B}{2}{2}{1})$ & 0.894 & 45\% \\
 $C(\orbh{B}{1}{2}{1} \;|\; \orbh{B}{2}{2}{1})$ & 0.605 & 30\% \\
 $C_{\text{$2$-part}}(X_1)$ & 1.500 & 75\% \\
 $C_{\text{$3$-part}}(X_1)$ & 2.394 & 60\% \\
 $C(X_1 \;|\; X_2)$ & 0.309 & 5.2\% \\
 $C(\orb{H}{5}{1}{s} \;|\; \orb{H}{6}{1}{s})$ & 0.042 & 2.1\% \\
 \hline
 \hline
\end{tabular}
\caption{Correlation measures for diborane(6). Relative values are related to the upper bounds. Labeling of localized orbitals corresponds to Figure \ref{fig:B2H6}.}
\label{table:B2H6}
\end{table}

Figure \ref{fig:B2H6} shows that our results fit well the established bonding picture of diborane(6) with two
bridging \mbox{B-H-B} bonds. Let us now discuss in detail how the ana\-ly\-sis of correlations leads to the clustering and to the bonding picture presented in Figure \ref{fig:B2H6}.
We will discuss only the bridging bonds, the core orbitals as well as terminal B-H bonds are well separated, i.e. not correlated with the rest. (This is confirmed by the weak correlation \mbox{$C(X_1 \cup X_2 \;|\; \rest)$} in Table \ref{table:B2H6}.)

First we consider the cluster $X_1$ containing sp$^3$ hybrid orbitals on B atoms and 1s orbital on the bridging H atom.
Because of the point group symmetry, the same results hold for cluster $X_2$. As can be seen in Table \ref{table:B2H6}, the correlation (entanglement) of $X_1$ with the remaining orbitals is very weak, only 8.6\% of the maximum value, which indicates that $X_1$ forms an independent three-center bond. 

However, to confirm this conclusion, we have to show that $X_1$ cannot be split further.
If we take separate pairs of orbitals from $X_1$
and measure the correlation with the remaining orbitals, we obtain significantly higher values, in particular 46\% and 53\% of the maximum values (see Table \ref{table:B2H6}), which justifies existence of the three-center bond.

The mutual information of pairs of orbitals within $X_1$ reach rather 
small relative values (45\%, 30\%, see Table \ref{table:B2H6}), however according to our numerical experience, 
even strong multicenter bonds typically yield low percentage, never 
approaching near the theoretical maxima.\cite{Szalay-2017}
Intuitive perspective suggests that the correlation of one orbital with the others can be thought of as a resource shared among the orbitals.
In other words, all the pairs inside $X_1$ cannot reach the maximum simultaneously,
bounded by entanglement monogamy.\cite{Osborne2006,Coffman2000} 
The formulation of an inequality bounding the mutual 
correlations inside orbital clusters still remains an open problem, to our best knowledge. 
The smaller value of the mutual information between sp$^3$ hybrid orbitals on B atoms reflects their larger internuclear distance.

We employ $k$-partitionability in order to 
quantify and benchmark the strength of the diborane(6)
three-center bonds (in terms of correlation).
As can be seen in Table \ref{table:B2H6},
$C_{\text{$2$-part}}(X_1)$ reaches 75\% of the upper bound and $C_{\text{$3$-part}}(X_1)$ 60\%, which points at a strong bond in $X_1$.

The very weak correlation (entanglement) of $X_1$ with the remaining orbitals  
also indicates that the state of the cluster $X_1$ is close to a pure state. 
Indeed, the eigenstate ana\-ly\-sis of the reduced density operator $\rho_{X_1}$ shows that there is the following two-electron eigenstate 
with a corresponding eigenvalue (probability) of $0.94$
\begin{align*}
	\vspace{-0.2cm}
	&\ket{\psi_{X_1}} = \\
		&+0.2146\ket{--\updownarrows}
		+0.4313\ket{-\updownarrows -}
		+0.2146\ket{\updownarrows--}\\
		&+0.3787\ket{-\downarrow\;\uparrow}
		-0.3787\ket{-\uparrow\;\downarrow}
		+0.2721\ket{\downarrow - \uparrow}\\
		&+0.3787\ket{\uparrow \; \downarrow -}
		-0.3787\ket{\downarrow \; \uparrow -}
		-0.2721\ket{\uparrow - \downarrow},
\end{align*}

\noindent
where the ordering of orbitals in a ket corresponds to \mbox{B$^1$sp$^3$}, \mbox{H$^5$1s}, and \mbox{B$^2$sp$^3$}.
Other eigenstates have probabilities below $0.01$. 
The principal two-electron eigenstate together with the 
above discussion on correlations imply that the three orbitals of $X_1$ form a three-center two-electron bond.
Note that the electron pair exhibits a preferred occupation on H atom, which is 
due to its higher electronegativity when compared to B, as we can see from the principal eigenstate. 
It is in agreement with the expectation values of particle-number-operators
($\bra{\psi} \hat{n}^{(i)}_{\text{el}} \ket{\psi}$)
which for \mbox{B$^1$sp$^3$}, \mbox{H$^5$1s} and \mbox{B$^2$sp$^3$} equal 0.53, 0,95 and 0.53, respectively.

As one can observe in Table \ref{table:B2H6}, 
the main source of correlation between $X_1$ and the remaining orbitals is the correlation with the other three-center bond, $X_2$. Specifically,
the correlation between two bridging H atoms is the strongest, which is caused by higher electron density on these atoms.

\subsection*{Diborane(4)}

\begin{figure}[!ht]
\centering
\includegraphics{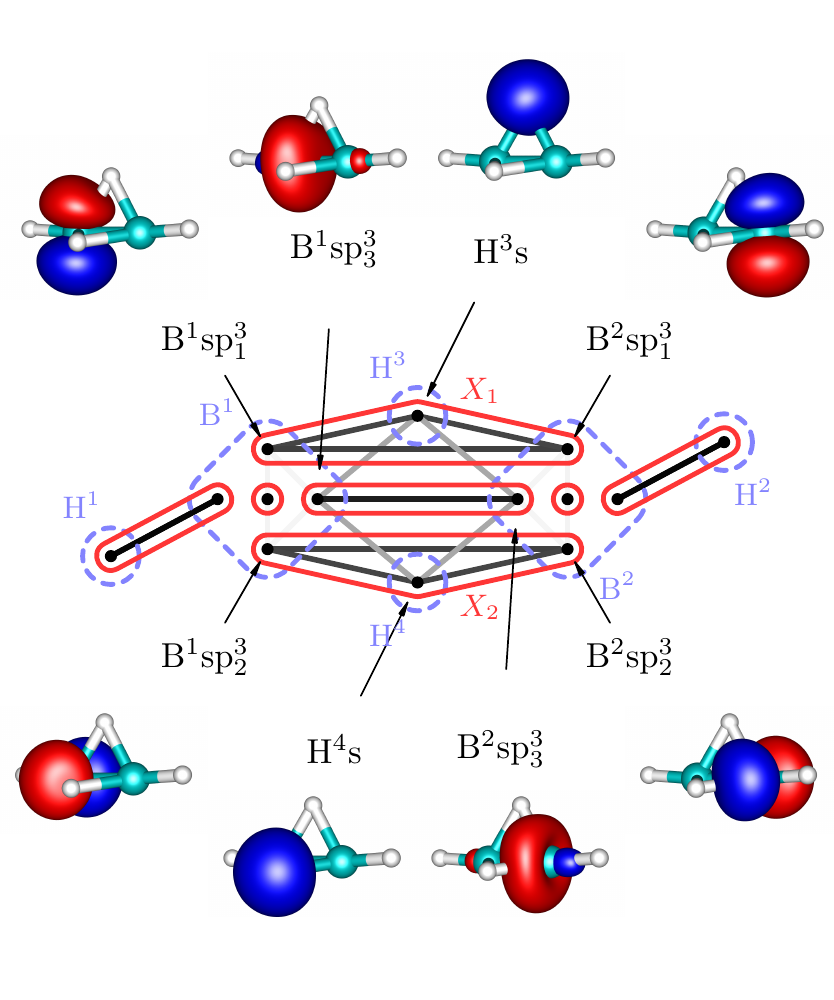}
\caption{Schematic view of diborane(4) with mutual information: 
	each dot represents a localized orbital, 
	dashed blue line encircles individual atoms,
	edges correspond to mutual information (plot shaded by a 
	logarithmic scale depending on strength) and 
red circles show how the orbitals group into clusters, i.e. independent bonds. Sorted values of the two-obital mutual information are plotted in Appendix \ref{appendix}.}
\label{fig:B2H4}
\end{figure}

\begin{table}[!ht]
\centering
\begin{tabular}{l c c}
 \hline
 \hline
 correlation & abs. value & rel. value \\ [0.5ex]
 \hline
 $C(X_1 \;|\; \rest)$ & 1.328 & 22\% \\
 $C(\orbh{B}{1}{2}{1} \;|\; \orb{H}{3}{1}{s})$ & 0.647 & 32\% \\
 $C(\orbh{B}{1}{2}{1} \;|\; \orbh{B}{2}{2}{1})$ & 0.701 & 35\% \\
 $C_{\text{$2$-part}}(X_1)$ & 1.388 & 69\% \\
 $C_{\text{$3$-part}}(X_1)$ & 2.089 & 52\% \\
 $C(X_3 \;|\; \rest)$ & 1.438 & 36\% \\
 $C(\orbh{B}{1}{2}{3} \;|\; \orbh{B}{2}{2}{3})$ & 1.245 & 62\% \\
 $C(\orbh{B}{1}{2}{3} \;|\; \orb{H}{3}{1}{s})$ & 0.130 & 6.5\% \\
 $C(X_1\cup X_2\cup X_3 \;|\; \rest)$ & 0.535 & 6.7\% \\
 $C(X_1 \;|\; X_2)$ & 0.066 & 1.1\% \\
 $C(X_1 \;|\; X_3)$ & 0.639 & 16\% \\ [1ex]
 \hline
 \hline
\end{tabular}
\caption{Correlation measures for diborane(4). Relative values are related to the upper bounds. Labeling of localized orbitals corresponds to Figure \ref{fig:B2H4}.}
\label{table:B2H4}
\end{table}

In case of diborane(4), the two terminal H atoms are missing and instead a direct covalent bond connecting both B atoms is present \cite{Chou2015}. This is also the picture resulting from our ana\-ly\-sis and depicted in Figure \ref{fig:B2H4}. In comparison to diborane(6), we have the similar three-orbital clusters $X_1$ and $X_2$, but also the two-orbital cluster $X_3$ containing sp$^3$ hybrid orbitals on B atoms and corresponding to the aforementioned \mbox{B-B} bond.

Considering the cluster $X_1$, one can observe in Table \ref{table:B2H4} that
it is more correlated with the remaining orbitals than in case of diborane(6). The value of $C(X_1 \;|\; \rest)$ is more than 
two times 
larger, 
but the picture of $X_1$ as a standalone chemical bond is still justifiable. 
Consequently, 
the reduced density operator $\rho_{X_1}$ is more mixed, with 
the principal eigenvalue $0.7803$. The remaining eigenstates share low
probabilities (below $0.081$), and therefore, the picture of $X_1$ as a standalone chemical bond is still a reasonable qualitative description. 
The principal eigenstate is again two-electron, i.e. electron-deficient,  and it has the following form
\begin{align*}
	\vspace{-0.2cm}
&\ket{\psi_{X_1}} =\\ 
	  &-0.2287\ket{--\updownarrows}
		-0.3671\ket{-\updownarrows -}
		-0.2287\ket{\updownarrows--}\\
		& +0.3823\ket{-\downarrow\;\uparrow}
		-0.3823\ket{-\uparrow\;\downarrow}
		-0.2967\ket{\downarrow - \uparrow}\\
		& +0.3823\ket{\uparrow \; \downarrow -}
		-0.3823\ket{\downarrow \; \uparrow -}
		+0.2967\ket{\uparrow - \downarrow}.
\end{align*}

\noindent
Similarly to diborane(6), higher electron density is on the bridging H atom, which is due to its higher electronegativity.

Comparing the two-orbital correlations inside $X_1$ with diborane(6) (Tables \ref{table:B2H6} and \ref{table:B2H4}), one can see a weaker correlation between sp$^3$ hybrid orbitals on B atoms and \mbox{H 1s} orbital, but a slightly stronger correlation between both B-atom-orbitals. This stronger correlation can be certainly assigned to a shorter distance of B atoms
(1.477\AA~vs. 1.784\AA).
Based on the values of $C_{\text{$2$-part}}(X_1)$ and $C_{\text{$3$-part}}(X_1)$, 
the covalent bond corresponding to the cluster $X_1$ is slightly weaker (in terms of correlation) than the same bond in diborane(6).

For the cluster $X_3 = \{\orbh{B}{1}{2}{3},\,\orbh{B}{2}{2}{3}\}$, correlation
with the remaining orbitals 
is stronger than for $X_1$ and $X_2$, which in turn weakens the internal two-orbital correlation.
The major contribution to the correlation of $X_3$ with the remaining orbitals 
originates from 
$C(\orbh{B}{1}{2}{3} \;|\; \orb{H}{3}{1}{s})$, which is still very weak compared to other correlations in the molecule (see Table \ref{table:B2H4}).

The overall correlation of the three bonding clusters $X_1$, $X_2$, and $X_3$ with the rest of the system is similarly 
weak as in diborane(6) 
so the considered bonding can be described independently of the rest of the molecule.


\subsection*{Beryllium complexes}

\begin{figure*}
\centering
\includegraphics{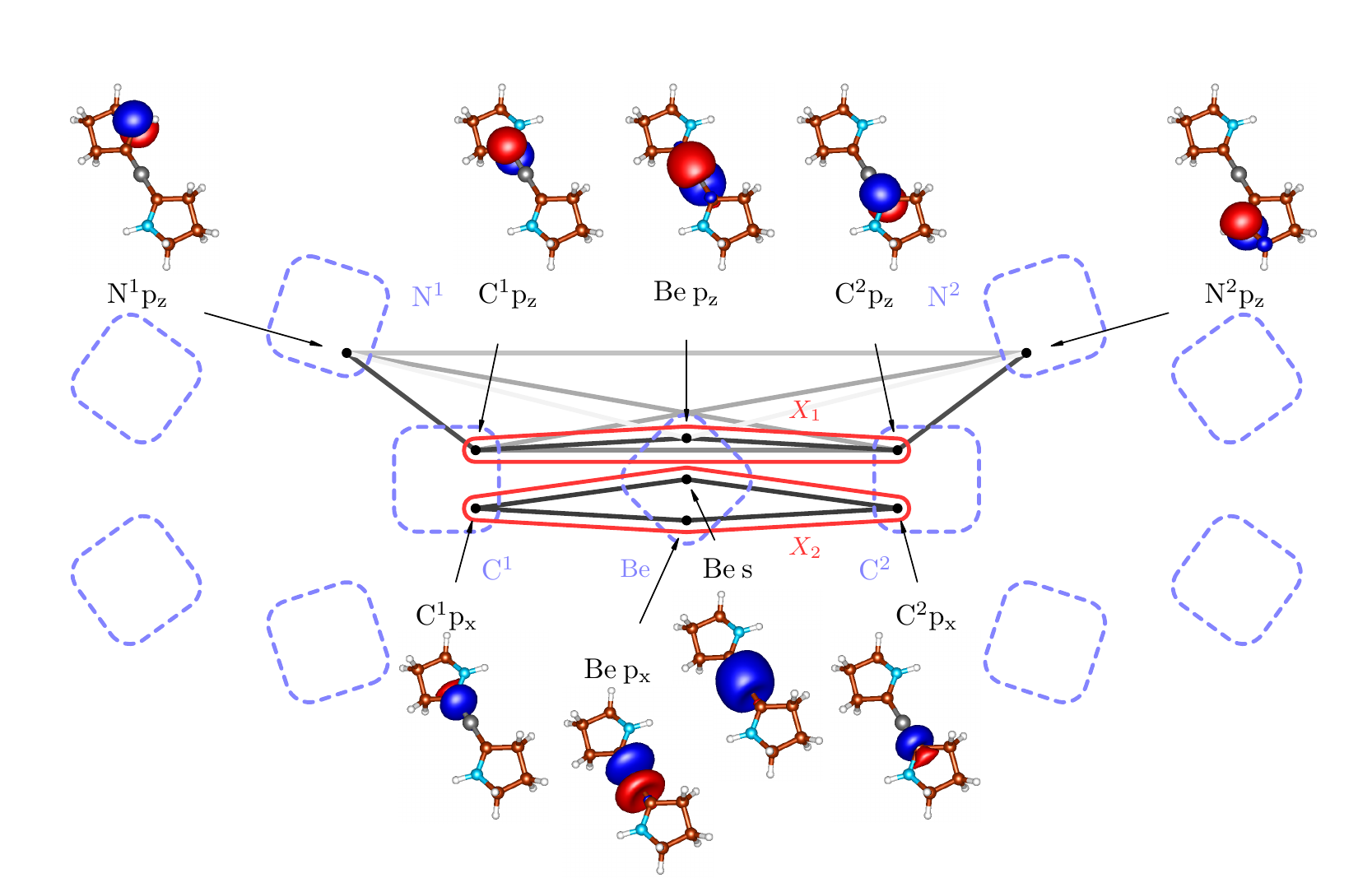}
\caption{Schematic view of Be(CAC)$_2$ with mutual information. In order to be consistent with Figures \ref{fig:B2H6} and \ref{fig:B2H4}, all the atoms are depicted (dashed blue line circles), even though only a subset of orbitals (black dots) formed the complete active space. Note that one ring is artificially flipped for clearer correlation picture.
	Sorted values of the two-obital mutual information are plotted in Appendix \ref{appendix}.}
\label{fig:CBeCv2}
\end{figure*}

In order to check how well our bonding picture of Be(CAC)$_2$ fits the one proposed by Arrowsmith \textit{et al.} \cite{Arrowsmith-2016}, we consider the clusters $X_1$ and $X_2$ from Figure \ref{fig:CBeCv2}. The cluster $X_1$ contains p$_\text{z}$ orbitals on C and Be atoms and corresponds to the suggested three-center two-electron $\pi$ bond, whereas the cluster $X_2$ contains \mbox{C p$_\text{x}$} orbitals and \mbox{Be s and p$_\text{x}$} orbitals and corresponds to the $\sigma$ bonds.

Let us start with $X_1$. In Table \ref{table:cbec01}, one can see that the correlation of $X_1$ with the remaining orbitals is higher than 30\% of the maximum value, which means that the picture of the three-orbital C-Be-C $\pi$ bond might be good as a qualitative description, but for a quantitatively adequate description, we might seek to include further orbitals into X$_1$, as shown below. This is also demonstrated by weaker pairwise correlations within $X_1$, especially $C(\orb{C}{1}{2}{p_z} \;|\; \orb{C}{2}{2}{p_z})$, than in the three-orbital bonds discussed above. The correlations of the internal pairs in $X_1$ with the remaining orbitals are large (61\% and 86\%) and clearly cannot be considered as standalone bonds. 

The inaccuracy of the picture of a standalone three-orbital bond is also demonstrated by the reduced density operator $\rho_{X_1}$ (see in Appendix \ref{appendix}), which is much more mixed than in previous cases. It has three dominant eigenvalues, instead of just one. The most significant, nevertheless, corresponds to the two-electron state, which is in agreement with the overall picture of the three-orbital two-electron bond.

\begin{table}[h!]
\centering
\begin{tabular}{l c c} 
 \hline
 \hline
 correlation & abs. value & rel. value \\ [0.5ex] 
 \hline
 $C(\orb{C}{1}{2}{p_z},\orb{C}{2}{2}{p_z} \;|\; \rest)$ & 3.424 & 86\% \\
  $C( \orb{C}{1}{2}{p_z},\orb{Be\,}{}{2}{p_z} \;|\; \rest)$ & 2.432 & 61\% \\
  $C(X_1 \;|\; \rest)$ & 2.032 & 34\% \\
  $C(\orb{C}{1}{2}{p_z} \;|\; \orb{C}{2}{2}{p_z})$ & 0.194 & 9.7\% \\
  $C(\orb{C}{1}{2}{p_z} \;|\; \orb{Be\,}{}{2}{p_z})$ & 0.681 & 34\% \\
  $C_{2-part}(X_1)$ & 1.153 & 58\% \\
  $C_{3-part}(X_1)$ & 1.834 & 46\% \\
  $C(X_1 \;|\; \orb{N}{1}{2}{p_z})$ & 0.915 & 46\% \\
  $C(X_1' \;|\; \rest)$ & 0.209 & 2.6\% \\
  $C(\orb{N}{1}{2}{p_z}, \orb{C}{1}{2}{p_z} \; | \; \mathrm{rest\; of}\;X_1')$ & 1.737  & 43\% \\
  $C(\orb{N}{1}{2}{p_z} \;|\; \orb{C}{1}{2}{p_z})$ & 0.560 & 28\% \\
  $C(X_2 \;|\; \rest)$ & 0.209 & 2.6\% \\
  $C(\orb{C}{1}{2}{p_x} \;|\; \orb{Be\,}{}{2}{s})$ & 0.765 & 38\% \\
  $C(\orb{C}{1}{2}{p_x} \;|\; \orb{Be\,}{}{2}{p_x}) $ & 0.771 & 39\% \\
  $C_{\text{$2$-part}}(X_2)$ & 1.936 & 97\% \\
  $C(\orb{C}{1}{2}{p_x} \;|\; \orb{Be\,}{}{}{sp_2})$ & 1.912 & 96\% \\
 \hline
 \hline
\end{tabular}
\caption{Correlation measures for Be(CAC)$_2$. Relative values are related to the upper bounds. Labeling of localized orbitals corresponds to Figure \ref{fig:CBeCv2}.}
\label{table:cbec01}
\end{table}

As can be seen in Figure \ref{fig:CBeCv2}, the strongest external correlation of $X_1$ is $C(X_1 \;|\; \orb{N}{1}{2}{p_z})$. 
It results from a conjugation of p$_\text{z}$ orbitals and has a stabilization effect. Notice that the N$^1$-C$^1$-Be-C$^2$-N$^2$ group of atoms form perfectly planar structure (the dihedral angle $\alpha_\text{N-C-C-N} = 179.97^{\circ}$) enabling an efficient overlap of all p$_{\text{z}}$ orbitals, which is necessary for the aforementioned conjugation.

\begin{figure}[!ht]
  \subfloat[Be(CAC)$_2$\label{str}]{%
    \includegraphics[width=4cm]{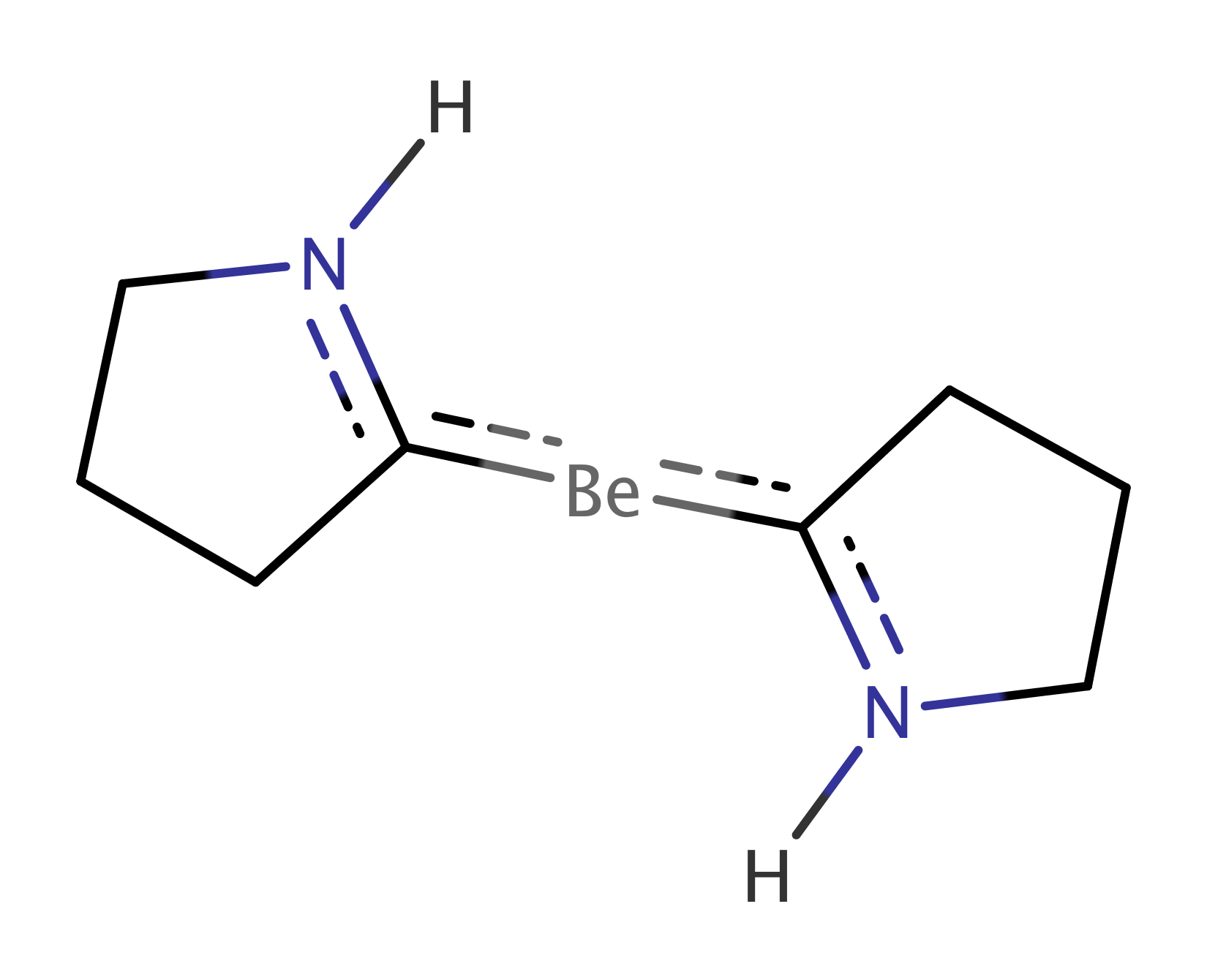}
  }
  \hfill
  \subfloat[Be(CAC)$_2^{2+}$\label{str2+}]{%
    \includegraphics[width=4cm]{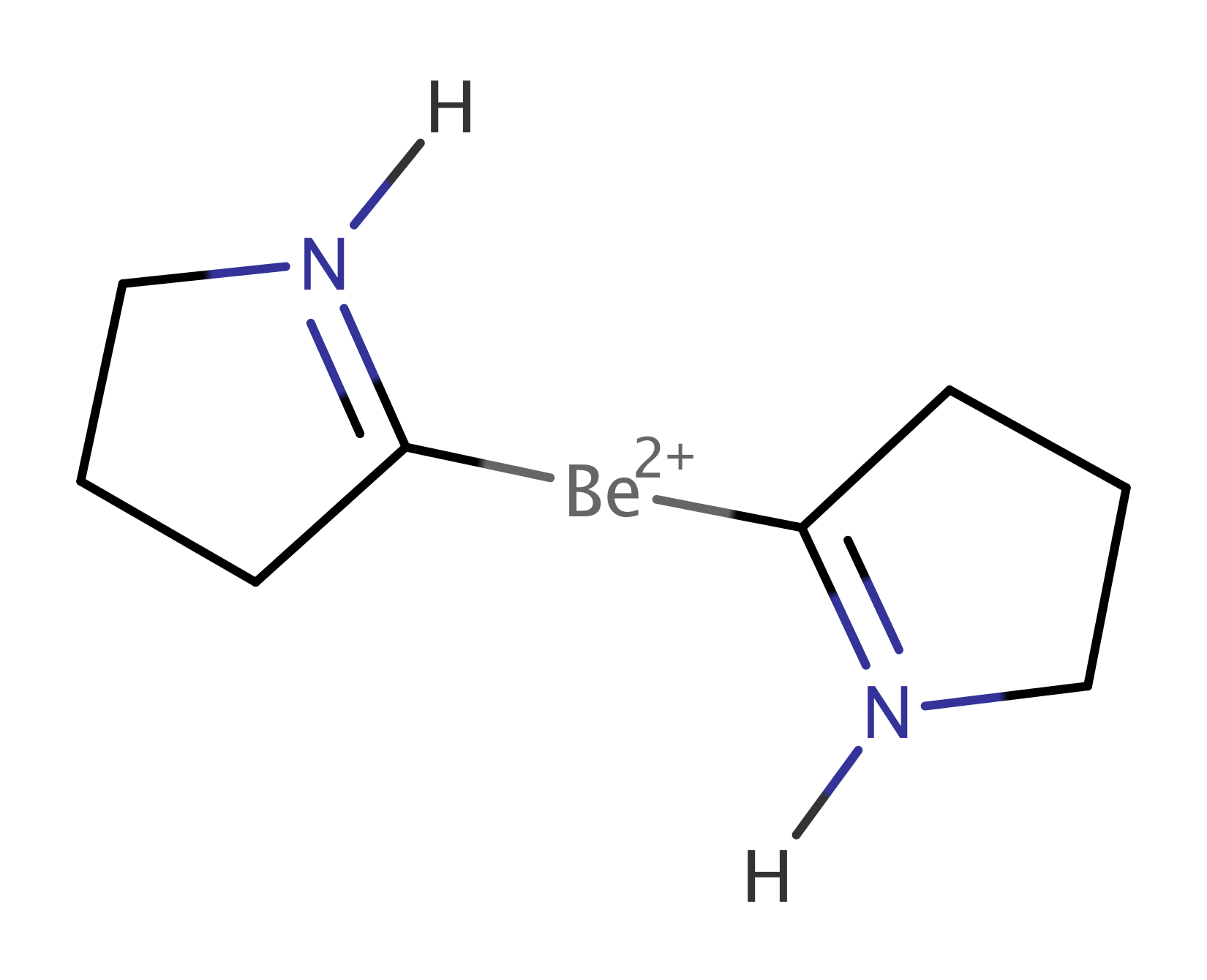}
  }
  \caption{Structures of studied beryllium complexes suggested by our correlation ana\-ly\-sis.}
\end{figure}

The more accurate bonding picture can thus be obtained by considering the enlarged cluster $X_1^{\prime}$
\begin{equation*}
	X_1 \longmapsto X_1'\equiv X_1\cup\{\orb{N}{1}{2}{p_z}, \orb{N}{2}{2}{p_z}\}.
\end{equation*}
\noindent
The corresponding structure of Be(CAC)$_2$ is depicted in Figure \ref{str}. The cluster $X_1^{\prime}$ is independent of the rest of the molecule. This follows from the negligible correlation of $X_1^{\prime}$ with the remaining orbitals (see Table \ref{table:cbec01}). Employing the standard notation, the $\pi$ electron bond can be denoted as $\Pi^{6}_{5}$, i.e. six-electron (two electrons from the Be atom and two from each N atom lone pair) five-center bond, which is confirmed by the particle number expectation value of $6.003$. 

\begin{figure*}
\centering
\includegraphics{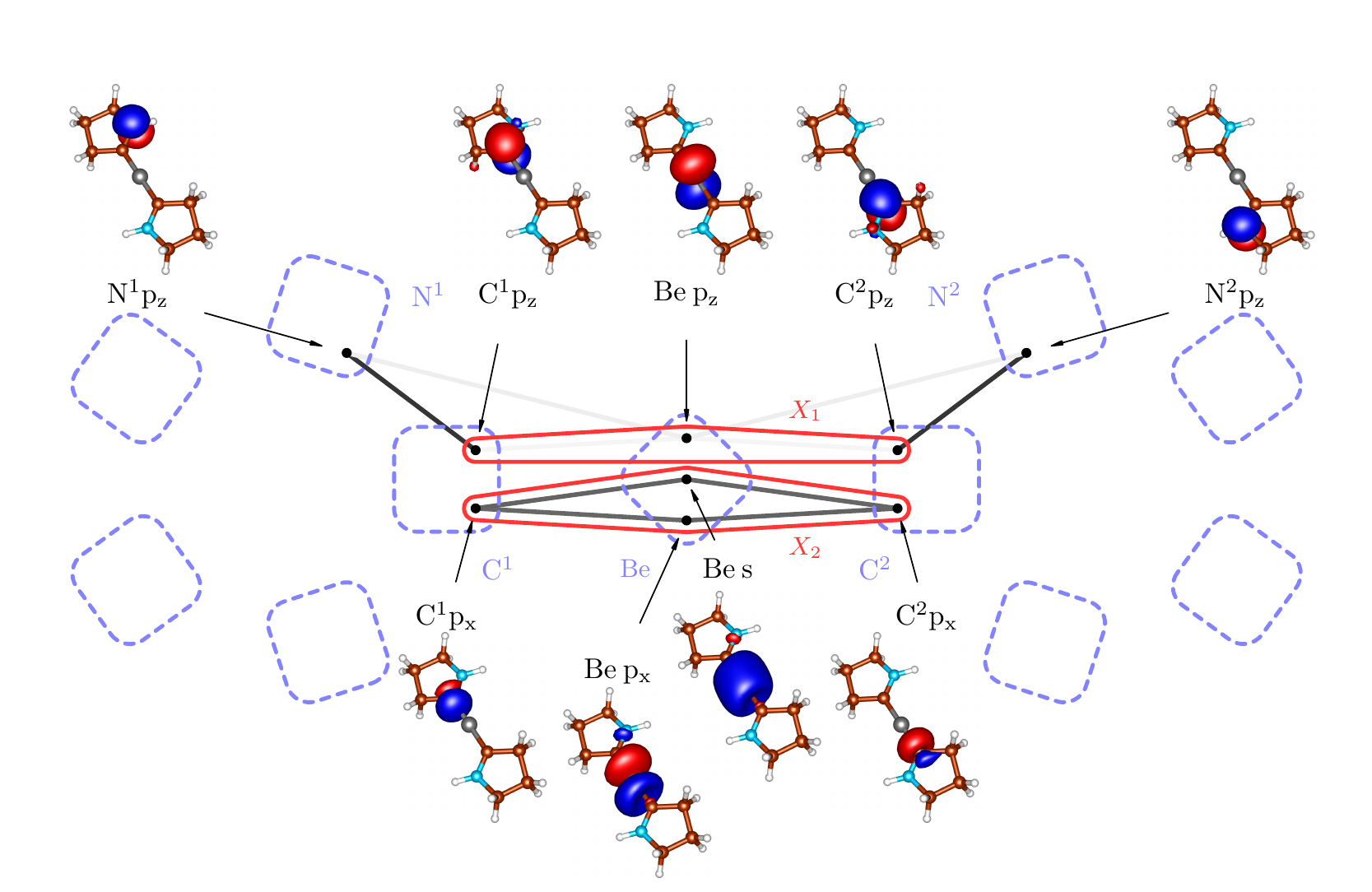}
\caption{Schematic view of [Be(CAC)$_2$]$^{2+}$ with mutual information. Note that one ring is artificially flipped for clearer correlation picture.}
\label{fig:dication}
\end{figure*}

In order to verify the suggested $\pi$ back-donation mechanism \cite{Arrowsmith-2016}, or in other words probe the local electronic configuration of the Be atom, we have also performed the correlation ana\-ly\-sis for the dication [Be(CAC)$_2$]$^{2+}$. As can be seen in Figure \ref{fig:dication}, the difference between the correlation picture of Be(CAC)$_2$ and [Be(CAC)$_2$]$^{2+}$ are almost missing correlations inside the cluster $X_1$, which is for example demonstrated by the negligible value of $C_{2-part}(X_1) = 0.093$. Also the reduced density operator $\rho_{X_1}$ is highly mixed and without the dominating two-electron eigenstate (see in Appendix \ref{appendix}). On the other hand, the correlations in the cluster $X_2$ remain practically unchanged. 

This is in agreement with the picture of Be atom having originally two electrons in the p$_z$ orbital. When the C-Be-C bond is formed, they are shared with C-atom p$_z$ orbitals through the back donation mechanism, as is depicted in Figure \ref{bonding_scheme}. These two $\pi$ electrons are missing in case of the dication and the aforementioned $\pi$ bond is clearly not formed.

Another feature of the correlation picture from Figure \ref{fig:dication} is that there are considerably stronger pairwise correlations between C and N-atom p$_z$ orbitals \mbox{[$C(\orb{N}{1}{2}{p_z} \;|\; \orb{C}{1}{2}{p_z}) = 1.691$]}. They are indeed of the strength of donor-acceptor bonds \cite{szilvasi_2015}. We thus assign double bonds between N and C atoms to the dication, as is depicted in Figure \ref{str2+}. The $\pi$ bonds are formed from the originally doubly filled N p$_\text{z}$ orbitals and empty C p$_\text{z}$ orbitals. The existence of these $\pi$ bonds is also confirmed by almost perfectly planar environment with the dihedral angle $\alpha_\text{H-N-C-Be} = 1.26^{\circ}$. Note that in Figure \ref{fig:dication}, we can see only the part of the double bond corresponding to the $\pi$ bond - the $\sigma$ bonds along the rings are excluded from the active space.

\begin{figure}
\centering
\vspace{-0.3cm}
\includegraphics{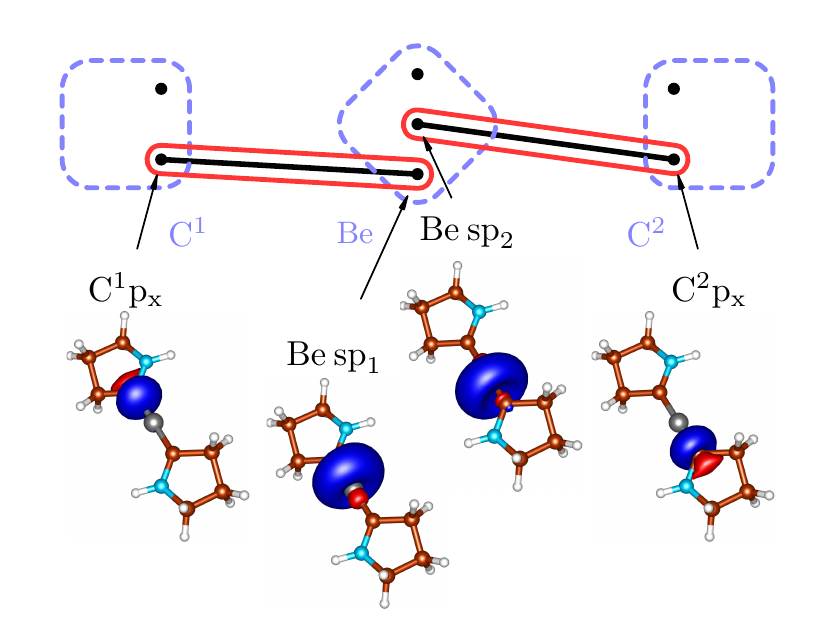}
\caption{Mutual information of the Be(CAC)$_2$ $\sigma$ bonding channel after the rotation of $\orb{Be\,}{}{2}{s}$ and $\orb{Be\,}{}{2}{p_x}$ orbitals.} 
\label{fig:cbecrotatedsmall}
\vspace{-0.1cm}
\end{figure}

Let us now turn to the $X_2$ cluster of Be(CAC)$_2$, i.e. to the $\sigma$ bonding. The correlation of $X_2$ with the remaining orbitals is insignificant, however splitting of the four-orbital cluster into two $\sigma$ bonds is not possible in this basis. It may therefore seem that the correct $\sigma$ bond is also multiorbital. It is, however, only the artefact of the atomic-like basis, which was used in order to directly compare with Ref.~\citen{Arrowsmith-2016}. By simple rotation of Be s and p$_x$ orbitals (forming the sp hybrids as in case of diboranes), one can form two independent $\sigma$ bonds, essentially without influencing the rest of the system, which can be seen in Figure \ref{fig:cbecrotatedsmall}. Note that in the correlation theory of the chemical bond, superposing orbitals is allowed if this does not affect their locality too much. So superposing orbitals on different atoms is usually forbidden, while doing the same on a given atom is allowed.

Last but not least, we would like to compare the strength of both contributions to the bonding of the \mbox{C-Be-C core,} namely the $\pi$ bond (clusters $X_1$ and $X_1^{\prime}$) and the $\sigma$ bond (cluster $X_2$). In the previous study \cite{Arrowsmith-2016}, it was shown by means of the 
energy decomposition ana\-ly\-sis combined with natural orbitals for chemical valence (EDA-NOCV) \cite{Krapp2007} that the $\pi$ bonding channel is considerably more energy stabilizing than the $\sigma$ channel. Let us now check the strength of both contributions by means of the correlations.

When using the more rough (three-orbital two-electron) description of the $\pi$ bond (cluster $X_1$) and by looking at Table \ref{table:cbec01}, one can see on the values of $C_{2-part}$ that $\sigma$ bonds are considerably stronger than the $\pi$ bond. 
The situation, however, dramatically changes  when we use more accurate conjugated description (cluster $X_1^{\prime}$).
Note that since we are not interested in a split dissecting N and C, 
(because the N-C bond on the ring is stabilized by another $\sigma$ bond, not visible on the plots),
we have to consider the p$_z$ orbitals together on N and C atoms of the same ring,
and calculate \mbox{$C_{2-part}( \orb{N}{1}{2}{p_z}, \orb{C}{1}{2}{p_z} \; | \; \orb{Be\,}{}{2}{p_z}\; | \;\orb{N}{2}{2}{p_z}, \orb{C}{2}{2}{p_z})$,} which turns out to be
$C(\orb{N}{1}{2}{p_z}, \orb{C}{1}{2}{p_z} \; | \; \mathrm{rest\; of}\;X_1')$.
Also note that in this paragraph we compared the absolute values of the correlation measures, because the clusters are of different sizes. 

In Table \ref{table:cbec01}, one can see that the more accurate description of the $\pi$ bond ($X_1^{\prime}$) makes it of a similar strength as the $\sigma$~bonds \mbox{(1.737 vs 1.936)}. We believe that our results describe the nature of a Be(CAC)$_2$ bonding reliably, especially because we have used the genuine multireference description unlike in Ref.~\citen{Arrowsmith-2016}, where the ana\-ly\-sis was based on the density functional theory (BP86 functional).
We would also like to note that we have studied slightly different system than in Ref.~\citen{Arrowsmith-2016}. In our case all substituents were replaced by hydrogen atoms. This, however, should not influence the electronic structure of the C-Be-C core.
Also note that using the s and p$_{\text{x}}$ orbitals on beryllium in $X_2$ was only for the purpose of comparison with the previous study\cite{Arrowsmith-2016}. For having a more physical picture, we should use the hybridized orbitals \mbox{(Figure \ref{fig:cbecrotatedsmall})}, by which $X_2$ consists of two simple covalent bonds. Then, in order to characterize the strength of the bond, we would wave to consider the Be sp$_1$ and sp$_2$ orbitals together and calculate 
$C_{2-part}( \orb{C}{1}{2}{p_x} \; | \; \text{Be\;sp}_1,\,\text{Be\;sp}_2 \; | \; \orb{C}{2}{2}{p_x}) = 1.935$. 
Nevertheless, since this value is nearly identical to $C_{2-part}(X_2)$, the conclusion is the same.

\section*{Conclusions}

In this article, we have reviewed the recently developed correlation theory of the chemical bond \cite{Szalay-2017} and applied it on molecules with multicenter electron-deficient bonds. We have demonstrated the usefulness of our methodology in characterizing molecular bonding properties by
fingerprints of correlations among individual orbitals which form these types of bonds.

We have verified the computational procedure on a textbook molecule with electron-deficient bonds, namely diborane(6), and further characterized bonding in diborane(4) and zero-valent complexes of beryllium with intricate bonding patterns. In all the cases, our results fit well with known bonding pictures or previous theoretical predictions.
We have therefore proved capabilities of our new method to reliably describe bonding in complex molecular systems.

In case of the Be(CAC)$_2$ molecule, we have also compared both contributions to the C-Be-C bonding ($\sigma$ and $\pi$), finding, in contrast to the previous study \cite{Arrowsmith-2016}, the $\sigma$ and $\pi$ contributions of a similar strength, in the sense of correlational quantities. We believe that our result is reliable and attribute the discrepancy with the previous study to the single reference description employed in Ref.~\citen{Arrowsmith-2016}, which may not be accurate enough 
in this multireference case.

Finally, we would like to note that, despite employing the DMRG method \cite{White-1992b, white_1993} for calculations of subsystem reduced density matrices, the theory presented in this article is general and other correlated methods
can in principle be employed as well \cite{Boguslawski-2015}. Especially in cases of large molecules with the electronic structure dominated by the dynamical correlation for which the DMRG description may be unnecessary and computationally prohibitive.

\section*{Acknowledgements}
This work has been supported by the \textit{Czech Science Foundation} (grants no. 16-12052S, 18-24563S, and 18-18940Y), 
\textit{Czech Ministry of Education, Youth and Sports} (project no. LTAUSA17033),
and the \textit{Hungarian-Czech Joint Research Project MTA/16/05}.
G.B., Sz.Sz. and {\"O}.L.~are supported by the National Research, 
Development and Innovation Fund of Hungary (NRDIFH) within the Researcher-initiated Research Program (project \mbox{Nr: NKFIH-K120569)} 
and the \textit{``Lend\"ulet'' Program} of the Hungarian Academy of Sciences (HAS).
Sz.Sz. and {\"O}.L. are supported by the Quantum Technology National Excellence Program 
(project \mbox{Nr: 2017-1.2.1-NKP-2017-00001)} of NRDIFH.
G.B. and Sz.Sz. are also supported by the \textit{``Bolyai'' Research Scholarship} of HAS.
{\"O}.L. also acknowledges financial support from the Alexander von Humboldt foundation.

\bibliographystyle{apsrev4-1}
\bibliography{paper_edeficitbonds}{}

\begin{thebibliography}{95}%
\makeatletter
\providecommand \@ifxundefined [1]{%
 \@ifx{#1\undefined}
}%
\providecommand \@ifnum [1]{%
 \ifnum #1\expandafter \@firstoftwo
 \else \expandafter \@secondoftwo
 \fi
}%
\providecommand \@ifx [1]{%
 \ifx #1\expandafter \@firstoftwo
 \else \expandafter \@secondoftwo
 \fi
}%
\providecommand \natexlab [1]{#1}%
\providecommand \enquote  [1]{``#1''}%
\providecommand \bibnamefont  [1]{#1}%
\providecommand \bibfnamefont [1]{#1}%
\providecommand \citenamefont [1]{#1}%
\providecommand \href@noop [0]{\@secondoftwo}%
\providecommand \href [0]{\begingroup \@sanitize@url \@href}%
\providecommand \@href[1]{\@@startlink{#1}\@@href}%
\providecommand \@@href[1]{\endgroup#1\@@endlink}%
\providecommand \@sanitize@url [0]{\catcode `\\12\catcode `\$12\catcode
  `\&12\catcode `\#12\catcode `\^12\catcode `\_12\catcode `\%12\relax}%
\providecommand \@@startlink[1]{}%
\providecommand \@@endlink[0]{}%
\providecommand \url  [0]{\begingroup\@sanitize@url \@url }%
\providecommand \@url [1]{\endgroup\@href {#1}{\urlprefix }}%
\providecommand \urlprefix  [0]{URL }%
\providecommand \Eprint [0]{\href }%
\providecommand \doibase [0]{http://dx.doi.org/}%
\providecommand \selectlanguage [0]{\@gobble}%
\providecommand \bibinfo  [0]{\@secondoftwo}%
\providecommand \bibfield  [0]{\@secondoftwo}%
\providecommand \translation [1]{[#1]}%
\providecommand \BibitemOpen [0]{}%
\providecommand \bibitemStop [0]{}%
\providecommand \bibitemNoStop [0]{.\EOS\space}%
\providecommand \EOS [0]{\spacefactor3000\relax}%
\providecommand \BibitemShut  [1]{\csname bibitem#1\endcsname}%
\let\auto@bib@innerbib\@empty
\bibitem [{\citenamefont {Legeza}\ and\ \citenamefont
  {S{\'o}lyom}(2003)}]{Legeza-2003b}%
  \BibitemOpen
  \bibfield  {author} {\bibinfo {author} {\bibfnamefont {{\"O}.}~\bibnamefont
  {Legeza}}\ and\ \bibinfo {author} {\bibfnamefont {J.}~\bibnamefont
  {S{\'o}lyom}},\ }\href {\doibase 10.1103/PhysRevB.68.195116} {\bibfield
  {journal} {\bibinfo  {journal} {Phys. Rev. B}\ }\textbf {\bibinfo {volume}
  {68}},\ \bibinfo {pages} {195116} (\bibinfo {year} {2003})}\BibitemShut
  {NoStop}%
\bibitem [{\citenamefont {Legeza}\ and\ \citenamefont
  {S{\'o}lyom}(2004)}]{Legeza-2004b}%
  \BibitemOpen
  \bibfield  {author} {\bibinfo {author} {\bibfnamefont {{\"O}.}~\bibnamefont
  {Legeza}}\ and\ \bibinfo {author} {\bibfnamefont {J.}~\bibnamefont
  {S{\'o}lyom}},\ }\href {\doibase 10.1103/PhysRevB.70.205118} {\bibfield
  {journal} {\bibinfo  {journal} {Phys. Rev. B}\ }\textbf {\bibinfo {volume}
  {70}},\ \bibinfo {pages} {205118} (\bibinfo {year} {2004})}\BibitemShut
  {NoStop}%
\bibitem [{\citenamefont {Huang}\ and\ \citenamefont {Kais}(2005)}]{Huang2005}%
  \BibitemOpen
  \bibfield  {author} {\bibinfo {author} {\bibfnamefont {Z.}~\bibnamefont
  {Huang}}\ and\ \bibinfo {author} {\bibfnamefont {S.}~\bibnamefont {Kais}},\
  }\href {\doibase 10.1016/j.cplett.2005.07.045} {\bibfield  {journal}
  {\bibinfo  {journal} {Chemical Physics Letters}\ }\textbf {\bibinfo {volume}
  {413}},\ \bibinfo {pages} {1} (\bibinfo {year} {2005})}\BibitemShut {NoStop}%
\bibitem [{\citenamefont {Rissler}\ \emph
  {et~al.}(2006{\natexlab{a}})\citenamefont {Rissler}, \citenamefont {Noack},\
  and\ \citenamefont {White}}]{rissler_2006}%
  \BibitemOpen
  \bibfield  {author} {\bibinfo {author} {\bibfnamefont {J.}~\bibnamefont
  {Rissler}}, \bibinfo {author} {\bibfnamefont {R.~M.}\ \bibnamefont {Noack}},
  \ and\ \bibinfo {author} {\bibfnamefont {S.~R.}\ \bibnamefont {White}},\
  }\href {\doibase 10.1016/j.chemphys.2005.10.018} {\bibfield  {journal}
  {\bibinfo  {journal} {Chem. Phys.}\ }\textbf {\bibinfo {volume} {323}},\
  \bibinfo {pages} {519} (\bibinfo {year} {2006}{\natexlab{a}})}\BibitemShut
  {NoStop}%
\bibitem [{\citenamefont {Pipek}\ and\ \citenamefont
  {Nagy}(2009)}]{Pipek-2009}%
  \BibitemOpen
  \bibfield  {author} {\bibinfo {author} {\bibfnamefont {J.}~\bibnamefont
  {Pipek}}\ and\ \bibinfo {author} {\bibfnamefont {I.}~\bibnamefont {Nagy}},\
  }\href {\doibase 10.1103/PhysRevA.79.052501} {\bibfield  {journal} {\bibinfo
  {journal} {Phys. Rev. A}\ }\textbf {\bibinfo {volume} {79}},\ \bibinfo
  {pages} {052501} (\bibinfo {year} {2009})}\BibitemShut {NoStop}%
\bibitem [{\citenamefont {Barcza}\ \emph {et~al.}(2011)\citenamefont {Barcza},
  \citenamefont {Legeza}, \citenamefont {Marti},\ and\ \citenamefont
  {Reiher}}]{Barcza-2011}%
  \BibitemOpen
  \bibfield  {author} {\bibinfo {author} {\bibfnamefont {G.}~\bibnamefont
  {Barcza}}, \bibinfo {author} {\bibfnamefont {{\"O}.}~\bibnamefont {Legeza}},
  \bibinfo {author} {\bibfnamefont {K.~H.}\ \bibnamefont {Marti}}, \ and\
  \bibinfo {author} {\bibfnamefont {M.}~\bibnamefont {Reiher}},\ }\href
  {\doibase 10.1103/PhysRevA.83.012508} {\bibfield  {journal} {\bibinfo
  {journal} {Phys. Rev. A}\ }\textbf {\bibinfo {volume} {83}},\ \bibinfo
  {pages} {012508} (\bibinfo {year} {2011})}\BibitemShut {NoStop}%
\bibitem [{\citenamefont {McKemmish}\ \emph {et~al.}(2011)\citenamefont
  {McKemmish}, \citenamefont {McKenzie}, \citenamefont {Hush},\ and\
  \citenamefont {Reimers}}]{McKemmish-2011}%
  \BibitemOpen
  \bibfield  {author} {\bibinfo {author} {\bibfnamefont {L.~K.}\ \bibnamefont
  {McKemmish}}, \bibinfo {author} {\bibfnamefont {R.~H.}\ \bibnamefont
  {McKenzie}}, \bibinfo {author} {\bibfnamefont {N.~S.}\ \bibnamefont {Hush}},
  \ and\ \bibinfo {author} {\bibfnamefont {J.~R.}\ \bibnamefont {Reimers}},\
  }\href {\doibase 10.1063/1.3671386} {\bibfield  {journal} {\bibinfo
  {journal} {The Journal of Chemical Physics}\ }\textbf {\bibinfo {volume}
  {135}},\ \bibinfo {eid} {244110} (\bibinfo {year} {2011}),\
  10.1063/1.3671386}\BibitemShut {NoStop}%
\bibitem [{\citenamefont {Boguslawski}\ \emph
  {et~al.}(2012{\natexlab{a}})\citenamefont {Boguslawski}, \citenamefont
  {Marti}, \citenamefont {Legeza},\ and\ \citenamefont
  {Reiher}}]{boguslawski_2012}%
  \BibitemOpen
  \bibfield  {author} {\bibinfo {author} {\bibfnamefont {K.}~\bibnamefont
  {Boguslawski}}, \bibinfo {author} {\bibfnamefont {K.~H.}\ \bibnamefont
  {Marti}}, \bibinfo {author} {\bibfnamefont {O.}~\bibnamefont {Legeza}}, \
  and\ \bibinfo {author} {\bibfnamefont {M.}~\bibnamefont {Reiher}},\
  }\href@noop {} {\bibfield  {journal} {\bibinfo  {journal} {J. Chem. Theory
  Comput.}\ }\textbf {\bibinfo {volume} {8}},\ \bibinfo {pages} {1970}
  (\bibinfo {year} {2012}{\natexlab{a}})}\BibitemShut {NoStop}%
\bibitem [{\citenamefont {Boguslawski}\ \emph
  {et~al.}(2012{\natexlab{b}})\citenamefont {Boguslawski}, \citenamefont
  {Tecmer}, \citenamefont {\"{O}rs Legeza},\ and\ \citenamefont
  {Reiher}}]{boguslawski_2012b}%
  \BibitemOpen
  \bibfield  {author} {\bibinfo {author} {\bibfnamefont {K.}~\bibnamefont
  {Boguslawski}}, \bibinfo {author} {\bibfnamefont {P.}~\bibnamefont {Tecmer}},
  \bibinfo {author} {\bibnamefont {\"{O}rs Legeza}}, \ and\ \bibinfo {author}
  {\bibfnamefont {M.}~\bibnamefont {Reiher}},\ }\href {\doibase
  10.1021/jz301319v} {\bibfield  {journal} {\bibinfo  {journal} {J. Phys. Chem.
  Lett.}\ }\textbf {\bibinfo {volume} {3}},\ \bibinfo {pages} {3129} (\bibinfo
  {year} {2012}{\natexlab{b}})}\BibitemShut {NoStop}%
\bibitem [{\citenamefont {Boguslawski}\ \emph {et~al.}(2013)\citenamefont
  {Boguslawski}, \citenamefont {Tecmer}, \citenamefont {Barcza}, \citenamefont
  {Legeza},\ and\ \citenamefont {Reiher}}]{Boguslawski-2013}%
  \BibitemOpen
  \bibfield  {author} {\bibinfo {author} {\bibfnamefont {K.}~\bibnamefont
  {Boguslawski}}, \bibinfo {author} {\bibfnamefont {P.}~\bibnamefont {Tecmer}},
  \bibinfo {author} {\bibfnamefont {G.}~\bibnamefont {Barcza}}, \bibinfo
  {author} {\bibfnamefont {{\"O}.}~\bibnamefont {Legeza}}, \ and\ \bibinfo
  {author} {\bibfnamefont {M.}~\bibnamefont {Reiher}},\ }\href {\doibase
  10.1021/ct400247p} {\bibfield  {journal} {\bibinfo  {journal} {Journal of
  Chemical Theory and Computation}\ }\textbf {\bibinfo {volume} {9}},\ \bibinfo
  {pages} {2959} (\bibinfo {year} {2013})}\BibitemShut {NoStop}%
\bibitem [{\citenamefont {Kurashige}\ \emph {et~al.}(2013)\citenamefont
  {Kurashige}, \citenamefont {Chan},\ and\ \citenamefont
  {Yanai}}]{Kurashige-2013}%
  \BibitemOpen
  \bibfield  {author} {\bibinfo {author} {\bibfnamefont {Y.}~\bibnamefont
  {Kurashige}}, \bibinfo {author} {\bibfnamefont {G.~K.-L.}\ \bibnamefont
  {Chan}}, \ and\ \bibinfo {author} {\bibfnamefont {T.}~\bibnamefont {Yanai}},\
  }\href {\doibase 10.1038/nchem.1677} {\bibfield  {journal} {\bibinfo
  {journal} {Nature Chemistry}\ }\textbf {\bibinfo {volume} {5}},\ \bibinfo
  {pages} {660} (\bibinfo {year} {2013})}\BibitemShut {NoStop}%
\bibitem [{\citenamefont {Fertitta}\ \emph {et~al.}(2014)\citenamefont
  {Fertitta}, \citenamefont {Paulus}, \citenamefont {Barcza},\ and\
  \citenamefont {Legeza}}]{Fertitta-2014}%
  \BibitemOpen
  \bibfield  {author} {\bibinfo {author} {\bibfnamefont {E.}~\bibnamefont
  {Fertitta}}, \bibinfo {author} {\bibfnamefont {B.}~\bibnamefont {Paulus}},
  \bibinfo {author} {\bibfnamefont {G.}~\bibnamefont {Barcza}}, \ and\ \bibinfo
  {author} {\bibfnamefont {{\"O}.}~\bibnamefont {Legeza}},\ }\href {\doibase
  10.1103/PhysRevB.90.245129} {\bibfield  {journal} {\bibinfo  {journal} {Phys.
  Rev. B}\ }\textbf {\bibinfo {volume} {90}},\ \bibinfo {pages} {245129}
  (\bibinfo {year} {2014})}\BibitemShut {NoStop}%
\bibitem [{\citenamefont {Duperrouzel}\ \emph {et~al.}(2015)\citenamefont
  {Duperrouzel}, \citenamefont {Tecmer}, \citenamefont {Boguslawski},
  \citenamefont {Barcza}, \citenamefont {Legeza},\ and\ \citenamefont
  {Ayers}}]{Duperrouzel-2014}%
  \BibitemOpen
  \bibfield  {author} {\bibinfo {author} {\bibfnamefont {C.}~\bibnamefont
  {Duperrouzel}}, \bibinfo {author} {\bibfnamefont {P.}~\bibnamefont {Tecmer}},
  \bibinfo {author} {\bibfnamefont {K.}~\bibnamefont {Boguslawski}}, \bibinfo
  {author} {\bibfnamefont {G.}~\bibnamefont {Barcza}}, \bibinfo {author}
  {\bibfnamefont {{\"O}.}~\bibnamefont {Legeza}}, \ and\ \bibinfo {author}
  {\bibfnamefont {P.~W.}\ \bibnamefont {Ayers}},\ }\href {\doibase
  10.1016/j.cplett.2015.01.005} {\bibfield  {journal} {\bibinfo  {journal}
  {Chemical Physics Letters}\ }\textbf {\bibinfo {volume} {621}},\ \bibinfo
  {pages} {160 } (\bibinfo {year} {2015})}\BibitemShut {NoStop}%
\bibitem [{\citenamefont {Murg}\ \emph
  {et~al.}(2015{\natexlab{a}})\citenamefont {Murg}, \citenamefont {Verstraete},
  \citenamefont {Schneider}, \citenamefont {Nagy},\ and\ \citenamefont
  {Legeza}}]{murg_2014}%
  \BibitemOpen
  \bibfield  {author} {\bibinfo {author} {\bibfnamefont {V.}~\bibnamefont
  {Murg}}, \bibinfo {author} {\bibfnamefont {F.}~\bibnamefont {Verstraete}},
  \bibinfo {author} {\bibfnamefont {R.}~\bibnamefont {Schneider}}, \bibinfo
  {author} {\bibfnamefont {P.~R.}\ \bibnamefont {Nagy}}, \ and\ \bibinfo
  {author} {\bibfnamefont {O.}~\bibnamefont {Legeza}},\ }\href@noop {}
  {\bibfield  {journal} {\bibinfo  {journal} {J. Chem. Theory Comput.}\
  }\textbf {\bibinfo {volume} {11}},\ \bibinfo {pages} {1027} (\bibinfo {year}
  {2015}{\natexlab{a}})}\BibitemShut {NoStop}%
\bibitem [{\citenamefont {Knecht}\ \emph {et~al.}(2014)\citenamefont {Knecht},
  \citenamefont {\"{O}rs Legeza},\ and\ \citenamefont {Reiher}}]{Knecht-2014}%
  \BibitemOpen
  \bibfield  {author} {\bibinfo {author} {\bibfnamefont {S.}~\bibnamefont
  {Knecht}}, \bibinfo {author} {\bibnamefont {\"{O}rs Legeza}}, \ and\ \bibinfo
  {author} {\bibfnamefont {M.}~\bibnamefont {Reiher}},\ }\href {\doibase
  10.1063/1.4862495} {\bibfield  {journal} {\bibinfo  {journal} {The Journal of
  Chemical Physics}\ }\textbf {\bibinfo {volume} {140}},\ \bibinfo {pages}
  {041101} (\bibinfo {year} {2014})}\BibitemShut {NoStop}%
\bibitem [{\citenamefont {Boguslawski}\ and\ \citenamefont
  {Tecmer}(2015)}]{Boguslawski-2015}%
  \BibitemOpen
  \bibfield  {author} {\bibinfo {author} {\bibfnamefont {K.}~\bibnamefont
  {Boguslawski}}\ and\ \bibinfo {author} {\bibfnamefont {P.}~\bibnamefont
  {Tecmer}},\ }\href {\doibase 10.1002/qua.24832} {\bibfield  {journal}
  {\bibinfo  {journal} {International Journal of Quantum Chemistry}\ }\textbf
  {\bibinfo {volume} {115}},\ \bibinfo {pages} {1289} (\bibinfo {year}
  {2015})}\BibitemShut {NoStop}%
\bibitem [{\citenamefont {{\relax Sz}alay}\ \emph {et~al.}(2015)\citenamefont
  {{\relax Sz}alay}, \citenamefont {Pfeffer}, \citenamefont {Murg},
  \citenamefont {Barcza}, \citenamefont {Verstraete}, \citenamefont
  {Schneider},\ and\ \citenamefont {Legeza}}]{Szalay-2015a}%
  \BibitemOpen
  \bibfield  {author} {\bibinfo {author} {\bibfnamefont {{\relax
  Sz}.}~\bibnamefont {{\relax Sz}alay}}, \bibinfo {author} {\bibfnamefont
  {M.}~\bibnamefont {Pfeffer}}, \bibinfo {author} {\bibfnamefont
  {V.}~\bibnamefont {Murg}}, \bibinfo {author} {\bibfnamefont {G.}~\bibnamefont
  {Barcza}}, \bibinfo {author} {\bibfnamefont {F.}~\bibnamefont {Verstraete}},
  \bibinfo {author} {\bibfnamefont {R.}~\bibnamefont {Schneider}}, \ and\
  \bibinfo {author} {\bibfnamefont {{\"O}.}~\bibnamefont {Legeza}},\ }\href
  {\doibase 10.1002/qua.24898} {\bibfield  {journal} {\bibinfo  {journal} {Int.
  J. Quantum Chem.}\ }\textbf {\bibinfo {volume} {115}},\ \bibinfo {pages}
  {1342} (\bibinfo {year} {2015})}\BibitemShut {NoStop}%
\bibitem [{\citenamefont {Freitag}\ \emph {et~al.}(2015)\citenamefont
  {Freitag}, \citenamefont {Knecht}, \citenamefont {Keller}, \citenamefont
  {Delcey}, \citenamefont {Aquilante}, \citenamefont {Bondo~Pedersen},
  \citenamefont {Lindh}, \citenamefont {Reiher},\ and\ \citenamefont
  {Gonzalez}}]{Freitag-2015}%
  \BibitemOpen
  \bibfield  {author} {\bibinfo {author} {\bibfnamefont {L.}~\bibnamefont
  {Freitag}}, \bibinfo {author} {\bibfnamefont {S.}~\bibnamefont {Knecht}},
  \bibinfo {author} {\bibfnamefont {S.~F.}\ \bibnamefont {Keller}}, \bibinfo
  {author} {\bibfnamefont {M.~G.}\ \bibnamefont {Delcey}}, \bibinfo {author}
  {\bibfnamefont {F.}~\bibnamefont {Aquilante}}, \bibinfo {author}
  {\bibfnamefont {T.}~\bibnamefont {Bondo~Pedersen}}, \bibinfo {author}
  {\bibfnamefont {R.}~\bibnamefont {Lindh}}, \bibinfo {author} {\bibfnamefont
  {M.}~\bibnamefont {Reiher}}, \ and\ \bibinfo {author} {\bibfnamefont
  {L.}~\bibnamefont {Gonzalez}},\ }\href {\doibase 10.1039/C4CP05278A}
  {\bibfield  {journal} {\bibinfo  {journal} {Phys. Chem. Chem. Phys.}\
  }\textbf {\bibinfo {volume} {17}},\ \bibinfo {pages} {14383} (\bibinfo {year}
  {2015})}\BibitemShut {NoStop}%
\bibitem [{\citenamefont {Zhao}\ \emph {et~al.}(2015)\citenamefont {Zhao},
  \citenamefont {Boguslawski}, \citenamefont {Tecmer}, \citenamefont
  {Duperrouzel}, \citenamefont {Barcza}, \citenamefont {Legeza},\ and\
  \citenamefont {Ayers}}]{Zhao-2015}%
  \BibitemOpen
  \bibfield  {author} {\bibinfo {author} {\bibfnamefont {Y.}~\bibnamefont
  {Zhao}}, \bibinfo {author} {\bibfnamefont {K.}~\bibnamefont {Boguslawski}},
  \bibinfo {author} {\bibfnamefont {P.}~\bibnamefont {Tecmer}}, \bibinfo
  {author} {\bibfnamefont {C.}~\bibnamefont {Duperrouzel}}, \bibinfo {author}
  {\bibfnamefont {G.}~\bibnamefont {Barcza}}, \bibinfo {author} {\bibfnamefont
  {{\"{O}}.}~\bibnamefont {Legeza}}, \ and\ \bibinfo {author} {\bibfnamefont
  {P.~W.}\ \bibnamefont {Ayers}},\ }\href {\doibase 10.1007/s00214-015-1726-3}
  {\bibfield  {journal} {\bibinfo  {journal} {Theor. Chem. Acc.}\ }\textbf
  {\bibinfo {volume} {134}},\ \bibinfo {pages} {120} (\bibinfo {year}
  {2015})}\BibitemShut {NoStop}%
\bibitem [{\citenamefont {{\relax Sz}ilv{\'a}si}\ \emph
  {et~al.}(2015)\citenamefont {{\relax Sz}ilv{\'a}si}, \citenamefont {Barcza},\
  and\ \citenamefont {Legeza}}]{Szilvasi-2015}%
  \BibitemOpen
  \bibfield  {author} {\bibinfo {author} {\bibfnamefont {T.}~\bibnamefont
  {{\relax Sz}ilv{\'a}si}}, \bibinfo {author} {\bibfnamefont {G.}~\bibnamefont
  {Barcza}}, \ and\ \bibinfo {author} {\bibfnamefont {{\"O}.}~\bibnamefont
  {Legeza}},\ }\href {http://arxiv.org/abs/1509.04241} {\bibfield  {journal}
  {\bibinfo  {journal} {arXiv [physics.chem-ph]}\ ,\ \bibinfo {pages}
  {1509.04241}} (\bibinfo {year} {2015})}\BibitemShut {NoStop}%
\bibitem [{\citenamefont {Molina-Esp{\'{\i}}ritu}\ \emph
  {et~al.}(2015)\citenamefont {Molina-Esp{\'{\i}}ritu}, \citenamefont
  {Esquivel}, \citenamefont {L{\'{o}}pez-Rosa},\ and\ \citenamefont
  {Dehesa}}]{Molina2015}%
  \BibitemOpen
  \bibfield  {author} {\bibinfo {author} {\bibfnamefont {M.}~\bibnamefont
  {Molina-Esp{\'{\i}}ritu}}, \bibinfo {author} {\bibfnamefont {R.~O.}\
  \bibnamefont {Esquivel}}, \bibinfo {author} {\bibfnamefont {S.}~\bibnamefont
  {L{\'{o}}pez-Rosa}}, \ and\ \bibinfo {author} {\bibfnamefont {J.~S.}\
  \bibnamefont {Dehesa}},\ }\href {\doibase 10.1021/acs.jctc.5b00390}
  {\bibfield  {journal} {\bibinfo  {journal} {Journal of Chemical Theory and
  Computation}\ }\textbf {\bibinfo {volume} {11}},\ \bibinfo {pages} {5144}
  (\bibinfo {year} {2015})}\BibitemShut {NoStop}%
\bibitem [{\citenamefont {Krumnow}\ \emph {et~al.}(2016)\citenamefont
  {Krumnow}, \citenamefont {Veis}, \citenamefont {Legeza},\ and\ \citenamefont
  {Eisert}}]{krumnow_2016}%
  \BibitemOpen
  \bibfield  {author} {\bibinfo {author} {\bibfnamefont {C.}~\bibnamefont
  {Krumnow}}, \bibinfo {author} {\bibfnamefont {L.}~\bibnamefont {Veis}},
  \bibinfo {author} {\bibfnamefont {O.}~\bibnamefont {Legeza}}, \ and\ \bibinfo
  {author} {\bibfnamefont {J.}~\bibnamefont {Eisert}},\ }\href@noop {}
  {\bibfield  {journal} {\bibinfo  {journal} {Phys. Rev. Lett.}\ }\textbf
  {\bibinfo {volume} {117}},\ \bibinfo {pages} {210402} (\bibinfo {year}
  {2016})}\BibitemShut {NoStop}%
\bibitem [{\citenamefont {Stein}\ and\ \citenamefont
  {Reiher}(2016)}]{stein_2016}%
  \BibitemOpen
  \bibfield  {author} {\bibinfo {author} {\bibfnamefont {C.~J.}\ \bibnamefont
  {Stein}}\ and\ \bibinfo {author} {\bibfnamefont {M.}~\bibnamefont {Reiher}},\
  }\href {\doibase 10.1021/acs.jctc.6b00156} {\bibfield  {journal} {\bibinfo
  {journal} {J. Chem. Theory Comput.}\ }\textbf {\bibinfo {volume} {12}},\
  \bibinfo {pages} {1760} (\bibinfo {year} {2016})}\BibitemShut {NoStop}%
\bibitem [{\citenamefont {Stein}\ and\ \citenamefont
  {Reiher}(2017)}]{stein_2017}%
  \BibitemOpen
  \bibfield  {author} {\bibinfo {author} {\bibfnamefont {C.}~\bibnamefont
  {Stein}}\ and\ \bibinfo {author} {\bibfnamefont {M.}~\bibnamefont {Reiher}},\
  }\href {\doibase 10.2533/chimia.2017.170} {\bibfield  {journal} {\bibinfo
  {journal} {Chimia}\ }\textbf {\bibinfo {volume} {71}},\ \bibinfo {pages}
  {170} (\bibinfo {year} {2017})}\BibitemShut {NoStop}%
\bibitem [{\citenamefont {Kovyrshin}\ and\ \citenamefont
  {Reiher}(2017)}]{kovyrshin_2017}%
  \BibitemOpen
  \bibfield  {author} {\bibinfo {author} {\bibfnamefont {A.}~\bibnamefont
  {Kovyrshin}}\ and\ \bibinfo {author} {\bibfnamefont {M.}~\bibnamefont
  {Reiher}},\ }\href@noop {} {\bibfield  {journal} {\bibinfo  {journal} {J.
  Chem. Phys.}\ }\textbf {\bibinfo {volume} {147}},\ \bibinfo {pages} {214111}
  (\bibinfo {year} {2017})}\BibitemShut {NoStop}%
\bibitem [{\citenamefont {{\relax Sz}alay}\ \emph {et~al.}(2017)\citenamefont
  {{\relax Sz}alay}, \citenamefont {Barcza}, \citenamefont {{\relax
  Sz}ilv{\'a}si}, \citenamefont {Veis},\ and\ \citenamefont
  {Legeza}}]{Szalay-2017}%
  \BibitemOpen
  \bibfield  {author} {\bibinfo {author} {\bibfnamefont {{\relax
  Sz}.}~\bibnamefont {{\relax Sz}alay}}, \bibinfo {author} {\bibfnamefont
  {G.}~\bibnamefont {Barcza}}, \bibinfo {author} {\bibfnamefont
  {T.}~\bibnamefont {{\relax Sz}ilv{\'a}si}}, \bibinfo {author} {\bibfnamefont
  {L.}~\bibnamefont {Veis}}, \ and\ \bibinfo {author} {\bibfnamefont
  {{\"O}.}~\bibnamefont {Legeza}},\ }\href {\doibase
  10.1038/s41598-017-02447-z} {\bibfield  {journal} {\bibinfo  {journal}
  {Scientific Reports}\ }\textbf {\bibinfo {volume} {7}},\ \bibinfo {pages}
  {2237} (\bibinfo {year} {2017})}\BibitemShut {NoStop}%
\bibitem [{\citenamefont {Stemmle}\ \emph {et~al.}(2018)\citenamefont
  {Stemmle}, \citenamefont {Paulus},\ and\ \citenamefont {\"{O}rs
  Legeza}}]{Stemmle2018}%
  \BibitemOpen
  \bibfield  {author} {\bibinfo {author} {\bibfnamefont {C.}~\bibnamefont
  {Stemmle}}, \bibinfo {author} {\bibfnamefont {B.}~\bibnamefont {Paulus}}, \
  and\ \bibinfo {author} {\bibnamefont {\"{O}rs Legeza}},\ }\href {\doibase
  10.1103/physreva.97.022505} {\bibfield  {journal} {\bibinfo  {journal}
  {Physical Review A}\ }\textbf {\bibinfo {volume} {97}} (\bibinfo {year}
  {2018}),\ 10.1103/physreva.97.022505}\BibitemShut {NoStop}%
\bibitem [{\citenamefont {Kurashige}\ and\ \citenamefont
  {Yanai}(2009)}]{Kurashige-2009}%
  \BibitemOpen
  \bibfield  {author} {\bibinfo {author} {\bibfnamefont {Y.}~\bibnamefont
  {Kurashige}}\ and\ \bibinfo {author} {\bibfnamefont {T.}~\bibnamefont
  {Yanai}},\ }\href {\doibase 10.1063/1.3152576} {\bibfield  {journal}
  {\bibinfo  {journal} {The Journal of Chemical Physics}\ }\textbf {\bibinfo
  {volume} {130}},\ \bibinfo {eid} {234114} (\bibinfo {year} {2009}),\
  10.1063/1.3152576}\BibitemShut {NoStop}%
\bibitem [{\citenamefont {Murg}\ \emph {et~al.}(2010)\citenamefont {Murg},
  \citenamefont {Verstraete}, \citenamefont {Legeza},\ and\ \citenamefont
  {Noack}}]{Murg-2010a}%
  \BibitemOpen
  \bibfield  {author} {\bibinfo {author} {\bibfnamefont {V.}~\bibnamefont
  {Murg}}, \bibinfo {author} {\bibfnamefont {F.}~\bibnamefont {Verstraete}},
  \bibinfo {author} {\bibfnamefont {O.}~\bibnamefont {Legeza}}, \ and\ \bibinfo
  {author} {\bibfnamefont {R.~M.}\ \bibnamefont {Noack}},\ }\href {\doibase
  10.1103/physrevb.82.205105} {\bibfield  {journal} {\bibinfo  {journal}
  {Physical Review B}\ }\textbf {\bibinfo {volume} {82}} (\bibinfo {year}
  {2010}),\ 10.1103/physrevb.82.205105}\BibitemShut {NoStop}%
\bibitem [{\citenamefont {Nakatani}\ and\ \citenamefont
  {Chan}(2013)}]{Nakatani-2013}%
  \BibitemOpen
  \bibfield  {author} {\bibinfo {author} {\bibfnamefont {N.}~\bibnamefont
  {Nakatani}}\ and\ \bibinfo {author} {\bibfnamefont {G.~K.-L.}\ \bibnamefont
  {Chan}},\ }\href {\doibase 10.1063/1.4798639} {\bibfield  {journal} {\bibinfo
   {journal} {The Journal of Chemical Physics}\ }\textbf {\bibinfo {volume}
  {138}},\ \bibinfo {pages} {134113} (\bibinfo {year} {2013})}\BibitemShut
  {NoStop}%
\bibitem [{\citenamefont {Chan}\ \emph {et~al.}(2016)\citenamefont {Chan},
  \citenamefont {Kesselman}, \citenamefont {Nakatani}, \citenamefont {Li},\
  and\ \citenamefont {White}}]{Chan-2015}%
  \BibitemOpen
  \bibfield  {author} {\bibinfo {author} {\bibfnamefont {G.~K.-L.}\
  \bibnamefont {Chan}}, \bibinfo {author} {\bibfnamefont {A.}~\bibnamefont
  {Kesselman}}, \bibinfo {author} {\bibfnamefont {N.}~\bibnamefont {Nakatani}},
  \bibinfo {author} {\bibfnamefont {Z.}~\bibnamefont {Li}}, \ and\ \bibinfo
  {author} {\bibfnamefont {S.~R.}\ \bibnamefont {White}},\ }\href {\doibase
  10.1063/1.4955108} {\bibfield  {journal} {\bibinfo  {journal} {The Journal of
  Chemical Physics}\ }\textbf {\bibinfo {volume} {145}},\ \bibinfo {pages}
  {014102} (\bibinfo {year} {2016})}\BibitemShut {NoStop}%
\bibitem [{\citenamefont {Keller}\ \emph {et~al.}(2015)\citenamefont {Keller},
  \citenamefont {Dolfi}, \citenamefont {Troyer},\ and\ \citenamefont
  {Reiher}}]{Reiher-MPO}%
  \BibitemOpen
  \bibfield  {author} {\bibinfo {author} {\bibfnamefont {S.}~\bibnamefont
  {Keller}}, \bibinfo {author} {\bibfnamefont {M.}~\bibnamefont {Dolfi}},
  \bibinfo {author} {\bibfnamefont {M.}~\bibnamefont {Troyer}}, \ and\ \bibinfo
  {author} {\bibfnamefont {M.}~\bibnamefont {Reiher}},\ }\href {\doibase
  10.1063/1.4939000} {\bibfield  {journal} {\bibinfo  {journal} {The Journal of
  Chemical Physics}\ }\textbf {\bibinfo {volume} {143}},\ \bibinfo {pages}
  {244118} (\bibinfo {year} {2015})}\BibitemShut {NoStop}%
\bibitem [{\citenamefont {Wouters}\ and\ \citenamefont
  {Van~Neck}(2014)}]{Wouters-2014e}%
  \BibitemOpen
  \bibfield  {author} {\bibinfo {author} {\bibfnamefont {S.}~\bibnamefont
  {Wouters}}\ and\ \bibinfo {author} {\bibfnamefont {D.}~\bibnamefont
  {Van~Neck}},\ }\href {\doibase 10.1140/epjd/e2014-50500-1} {\bibfield
  {journal} {\bibinfo  {journal} {The European Physical Journal D}\ }\textbf
  {\bibinfo {volume} {68}},\ \bibinfo {eid} {272} (\bibinfo {year} {2014}),\
  10.1140/epjd/e2014-50500-1}\BibitemShut {NoStop}%
\bibitem [{\citenamefont {Gunst}\ \emph {et~al.}(2018)\citenamefont {Gunst},
  \citenamefont {Verstraete}, \citenamefont {Wouters}, \citenamefont {\"{O}rs
  Legeza},\ and\ \citenamefont {Neck}}]{Gunst-2018}%
  \BibitemOpen
  \bibfield  {author} {\bibinfo {author} {\bibfnamefont {K.}~\bibnamefont
  {Gunst}}, \bibinfo {author} {\bibfnamefont {F.}~\bibnamefont {Verstraete}},
  \bibinfo {author} {\bibfnamefont {S.}~\bibnamefont {Wouters}}, \bibinfo
  {author} {\bibnamefont {\"{O}rs Legeza}}, \ and\ \bibinfo {author}
  {\bibfnamefont {D.~V.}\ \bibnamefont {Neck}},\ }\href {\doibase
  10.1021/acs.jctc.8b00098} {\bibfield  {journal} {\bibinfo  {journal} {Journal
  of Chemical Theory and Computation}\ }\textbf {\bibinfo {volume} {14}},\
  \bibinfo {pages} {2026} (\bibinfo {year} {2018})}\BibitemShut {NoStop}%
\bibitem [{\citenamefont {White}(1992)}]{White-1992b}%
  \BibitemOpen
  \bibfield  {author} {\bibinfo {author} {\bibfnamefont {S.~R.}\ \bibnamefont
  {White}},\ }\href {\doibase 10.1103/PhysRevLett.69.2863} {\bibfield
  {journal} {\bibinfo  {journal} {Phys. Rev. Lett.}\ }\textbf {\bibinfo
  {volume} {69}},\ \bibinfo {pages} {2863} (\bibinfo {year}
  {1992})}\BibitemShut {NoStop}%
\bibitem [{\citenamefont {White}(1993)}]{white_1993}%
  \BibitemOpen
  \bibfield  {author} {\bibinfo {author} {\bibfnamefont {S.~R.}\ \bibnamefont
  {White}},\ }\href@noop {} {\bibfield  {journal} {\bibinfo  {journal} {Phys.
  Rev. B}\ }\textbf {\bibinfo {volume} {48}},\ \bibinfo {pages} {10345}
  (\bibinfo {year} {1993})}\BibitemShut {NoStop}%
\bibitem [{\citenamefont {Schollw\"ock}(2011)}]{schollwock_2011}%
  \BibitemOpen
  \bibfield  {author} {\bibinfo {author} {\bibfnamefont {U.}~\bibnamefont
  {Schollw\"ock}},\ }\href@noop {} {\bibfield  {journal} {\bibinfo  {journal}
  {Ann. Phys.}\ }\textbf {\bibinfo {volume} {326}},\ \bibinfo {pages} {96 }
  (\bibinfo {year} {2011})}\BibitemShut {NoStop}%
\bibitem [{\citenamefont {Faulstich}\ \emph {et~al.}(2018)\citenamefont
  {Faulstich}, \citenamefont {Máté}, \citenamefont {Laestadius},
  \citenamefont {Csirik}, \citenamefont {Veis}, \citenamefont {Antalik},
  \citenamefont {Brabec}, \citenamefont {Schneider}, \citenamefont {Pittner},
  \citenamefont {Kvaal},\ and\ \citenamefont {Örs Legeza}}]{Faulstich2018}%
  \BibitemOpen
  \bibfield  {author} {\bibinfo {author} {\bibfnamefont {F.~M.}\ \bibnamefont
  {Faulstich}}, \bibinfo {author} {\bibfnamefont {M.}~\bibnamefont {Máté}},
  \bibinfo {author} {\bibfnamefont {A.}~\bibnamefont {Laestadius}}, \bibinfo
  {author} {\bibfnamefont {M.~A.}\ \bibnamefont {Csirik}}, \bibinfo {author}
  {\bibfnamefont {L.}~\bibnamefont {Veis}}, \bibinfo {author} {\bibfnamefont
  {A.}~\bibnamefont {Antalik}}, \bibinfo {author} {\bibfnamefont
  {J.}~\bibnamefont {Brabec}}, \bibinfo {author} {\bibfnamefont
  {R.}~\bibnamefont {Schneider}}, \bibinfo {author} {\bibfnamefont
  {J.}~\bibnamefont {Pittner}}, \bibinfo {author} {\bibfnamefont
  {S.}~\bibnamefont {Kvaal}}, \ and\ \bibinfo {author} {\bibnamefont {Örs
  Legeza}},\ }\href@noop {} {\enquote {\bibinfo {title} {Numerical and
  theoretical aspects of the dmrg-tcc method exemplified by the nitrogen
  dimer},}\ } (\bibinfo {year} {2018}),\ \Eprint
  {http://arxiv.org/abs/arXiv:1809.07732} {arXiv:1809.07732} \BibitemShut
  {NoStop}%
\bibitem [{Note1()}]{Note1}%
  \BibitemOpen
  \bibinfo {note} {The number in parentheses denotes the number of hydrogen
  atoms.}\BibitemShut {Stop}%
\bibitem [{\citenamefont {Chou}\ \emph {et~al.}(2015)\citenamefont {Chou},
  \citenamefont {Lo}, \citenamefont {Peng}, \citenamefont {Lin}, \citenamefont
  {Lu}, \citenamefont {Cheng},\ and\ \citenamefont {Ogilvie}}]{Chou2015}%
  \BibitemOpen
  \bibfield  {author} {\bibinfo {author} {\bibfnamefont {S.-L.}\ \bibnamefont
  {Chou}}, \bibinfo {author} {\bibfnamefont {J.-I.}\ \bibnamefont {Lo}},
  \bibinfo {author} {\bibfnamefont {Y.-C.}\ \bibnamefont {Peng}}, \bibinfo
  {author} {\bibfnamefont {M.-Y.}\ \bibnamefont {Lin}}, \bibinfo {author}
  {\bibfnamefont {H.-C.}\ \bibnamefont {Lu}}, \bibinfo {author} {\bibfnamefont
  {B.-M.}\ \bibnamefont {Cheng}}, \ and\ \bibinfo {author} {\bibfnamefont
  {J.~F.}\ \bibnamefont {Ogilvie}},\ }\href {\doibase 10.1039/c5sc02586a}
  {\bibfield  {journal} {\bibinfo  {journal} {Chemical Science}\ }\textbf
  {\bibinfo {volume} {6}},\ \bibinfo {pages} {6872} (\bibinfo {year}
  {2015})}\BibitemShut {NoStop}%
\bibitem [{\citenamefont {Arrowsmith}\ \emph {et~al.}(2016)\citenamefont
  {Arrowsmith}, \citenamefont {Braunschweig}, \citenamefont {Celik},
  \citenamefont {Dellermann}, \citenamefont {Dewhurst}, \citenamefont {Ewing},
  \citenamefont {Hammond}, \citenamefont {Kramer}, \citenamefont
  {Krummenacher}, \citenamefont {Mies}, \citenamefont {Radacki},\ and\
  \citenamefont {Schuster}}]{Arrowsmith-2016}%
  \BibitemOpen
  \bibfield  {author} {\bibinfo {author} {\bibfnamefont {M.}~\bibnamefont
  {Arrowsmith}}, \bibinfo {author} {\bibfnamefont {H.}~\bibnamefont
  {Braunschweig}}, \bibinfo {author} {\bibfnamefont {M.~A.}\ \bibnamefont
  {Celik}}, \bibinfo {author} {\bibfnamefont {T.}~\bibnamefont {Dellermann}},
  \bibinfo {author} {\bibfnamefont {R.~D.}\ \bibnamefont {Dewhurst}}, \bibinfo
  {author} {\bibfnamefont {W.~C.}\ \bibnamefont {Ewing}}, \bibinfo {author}
  {\bibfnamefont {K.}~\bibnamefont {Hammond}}, \bibinfo {author} {\bibfnamefont
  {T.}~\bibnamefont {Kramer}}, \bibinfo {author} {\bibfnamefont
  {I.}~\bibnamefont {Krummenacher}}, \bibinfo {author} {\bibfnamefont
  {J.}~\bibnamefont {Mies}}, \bibinfo {author} {\bibfnamefont {K.}~\bibnamefont
  {Radacki}}, \ and\ \bibinfo {author} {\bibfnamefont {J.~K.}\ \bibnamefont
  {Schuster}},\ }\href {\doibase 10.1038/nchem.2542} {\bibfield  {journal}
  {\bibinfo  {journal} {Nature Chemistry}\ }\textbf {\bibinfo {volume} {8}},\
  \bibinfo {pages} {890} (\bibinfo {year} {2016})}\BibitemShut {NoStop}%
\bibitem [{\citenamefont {Brabec}\ \emph {et~al.}(2018)\citenamefont {Brabec},
  \citenamefont {Lang}, \citenamefont {Saitow}, \citenamefont {Pittner},
  \citenamefont {Neese},\ and\ \citenamefont {Demel}}]{Brabec2018}%
  \BibitemOpen
  \bibfield  {author} {\bibinfo {author} {\bibfnamefont {J.}~\bibnamefont
  {Brabec}}, \bibinfo {author} {\bibfnamefont {J.}~\bibnamefont {Lang}},
  \bibinfo {author} {\bibfnamefont {M.}~\bibnamefont {Saitow}}, \bibinfo
  {author} {\bibfnamefont {J.}~\bibnamefont {Pittner}}, \bibinfo {author}
  {\bibfnamefont {F.}~\bibnamefont {Neese}}, \ and\ \bibinfo {author}
  {\bibfnamefont {O.}~\bibnamefont {Demel}},\ }\href {\doibase
  10.1021/acs.jctc.7b01184} {\bibfield  {journal} {\bibinfo  {journal} {J.
  Chem. Theor. Comput.}\ }\textbf {\bibinfo {volume} {14}},\ \bibinfo {pages}
  {1370} (\bibinfo {year} {2018})}\BibitemShut {NoStop}%
\bibitem [{\citenamefont {Longuet-Higgins}(1946)}]{LonguetHiggins1946}%
  \BibitemOpen
  \bibfield  {author} {\bibinfo {author} {\bibfnamefont {H.~C.}\ \bibnamefont
  {Longuet-Higgins}},\ }\href {\doibase 10.1039/jr9460000139} {\bibfield
  {journal} {\bibinfo  {journal} {Journal of the Chemical Society (Resumed)}\
  ,\ \bibinfo {pages} {139}} (\bibinfo {year} {1946})}\BibitemShut {NoStop}%
\bibitem [{\citenamefont {Longuet-Higgins}\ and\ \citenamefont
  {Bell}(1943)}]{LonguetHiggins1943}%
  \BibitemOpen
  \bibfield  {author} {\bibinfo {author} {\bibfnamefont {H.~C.}\ \bibnamefont
  {Longuet-Higgins}}\ and\ \bibinfo {author} {\bibfnamefont {R.~P.}\
  \bibnamefont {Bell}},\ }\href {\doibase 10.1039/jr9430000250} {\bibfield
  {journal} {\bibinfo  {journal} {Journal of the Chemical Society (Resumed)}\
  ,\ \bibinfo {pages} {250}} (\bibinfo {year} {1943})}\BibitemShut {NoStop}%
\bibitem [{\citenamefont {Eberhardt}\ \emph {et~al.}(1954)\citenamefont
  {Eberhardt}, \citenamefont {Crawford},\ and\ \citenamefont
  {Lipscomb}}]{Eberhardt1954}%
  \BibitemOpen
  \bibfield  {author} {\bibinfo {author} {\bibfnamefont {W.~H.}\ \bibnamefont
  {Eberhardt}}, \bibinfo {author} {\bibfnamefont {B.}~\bibnamefont {Crawford}},
  \ and\ \bibinfo {author} {\bibfnamefont {W.~N.}\ \bibnamefont {Lipscomb}},\
  }\href {\doibase 10.1063/1.1740320} {\bibfield  {journal} {\bibinfo
  {journal} {The Journal of Chemical Physics}\ }\textbf {\bibinfo {volume}
  {22}},\ \bibinfo {pages} {989} (\bibinfo {year} {1954})}\BibitemShut
  {NoStop}%
\bibitem [{\citenamefont {Lammertsma}\ and\ \citenamefont
  {Ohwada}(1996)}]{Lammertsma1996}%
  \BibitemOpen
  \bibfield  {author} {\bibinfo {author} {\bibfnamefont {K.}~\bibnamefont
  {Lammertsma}}\ and\ \bibinfo {author} {\bibfnamefont {T.}~\bibnamefont
  {Ohwada}},\ }\href {\doibase 10.1021/ja960004x} {\bibfield  {journal}
  {\bibinfo  {journal} {Journal of the American Chemical Society}\ }\textbf
  {\bibinfo {volume} {118}},\ \bibinfo {pages} {7247} (\bibinfo {year}
  {1996})}\BibitemShut {NoStop}%
\bibitem [{\citenamefont {Lipscomb}(1973)}]{Lipscomb1973}%
  \BibitemOpen
  \bibfield  {author} {\bibinfo {author} {\bibfnamefont {W.~N.}\ \bibnamefont
  {Lipscomb}},\ }\href {\doibase 10.1021/ar50068a001} {\bibfield  {journal}
  {\bibinfo  {journal} {Accounts of Chemical Research}\ }\textbf {\bibinfo
  {volume} {6}},\ \bibinfo {pages} {257} (\bibinfo {year} {1973})}\BibitemShut
  {NoStop}%
\bibitem [{\citenamefont {Neeve}\ \emph {et~al.}(2016)\citenamefont {Neeve},
  \citenamefont {Geier}, \citenamefont {Mkhalid}, \citenamefont {Westcott},\
  and\ \citenamefont {Marder}}]{Neeve2016}%
  \BibitemOpen
  \bibfield  {author} {\bibinfo {author} {\bibfnamefont {E.~C.}\ \bibnamefont
  {Neeve}}, \bibinfo {author} {\bibfnamefont {S.~J.}\ \bibnamefont {Geier}},
  \bibinfo {author} {\bibfnamefont {I.~A.~I.}\ \bibnamefont {Mkhalid}},
  \bibinfo {author} {\bibfnamefont {S.~A.}\ \bibnamefont {Westcott}}, \ and\
  \bibinfo {author} {\bibfnamefont {T.~B.}\ \bibnamefont {Marder}},\ }\href
  {\doibase 10.1021/acs.chemrev.6b00193} {\bibfield  {journal} {\bibinfo
  {journal} {Chemical Reviews}\ }\textbf {\bibinfo {volume} {116}},\ \bibinfo
  {pages} {9091} (\bibinfo {year} {2016})}\BibitemShut {NoStop}%
\bibitem [{\citenamefont {Vincent}\ and\ \citenamefont
  {Schaefer}(1981)}]{Vincent1981}%
  \BibitemOpen
  \bibfield  {author} {\bibinfo {author} {\bibfnamefont {M.~A.}\ \bibnamefont
  {Vincent}}\ and\ \bibinfo {author} {\bibfnamefont {H.~F.}\ \bibnamefont
  {Schaefer}},\ }\href {\doibase 10.1021/ja00409a008} {\bibfield  {journal}
  {\bibinfo  {journal} {Journal of the American Chemical Society}\ }\textbf
  {\bibinfo {volume} {103}},\ \bibinfo {pages} {5677} (\bibinfo {year}
  {1981})}\BibitemShut {NoStop}%
\bibitem [{\citenamefont {Mohr}\ and\ \citenamefont
  {Lipscomb}(1986)}]{Mohr1986}%
  \BibitemOpen
  \bibfield  {author} {\bibinfo {author} {\bibfnamefont {R.~R.}\ \bibnamefont
  {Mohr}}\ and\ \bibinfo {author} {\bibfnamefont {W.~N.}\ \bibnamefont
  {Lipscomb}},\ }\href {\doibase 10.1021/ic00227a033} {\bibfield  {journal}
  {\bibinfo  {journal} {Inorganic Chemistry}\ }\textbf {\bibinfo {volume}
  {25}},\ \bibinfo {pages} {1053} (\bibinfo {year} {1986})}\BibitemShut
  {NoStop}%
\bibitem [{\citenamefont {Curtiss}\ and\ \citenamefont
  {Pople}(1989{\natexlab{a}})}]{Curtiss1989}%
  \BibitemOpen
  \bibfield  {author} {\bibinfo {author} {\bibfnamefont {L.~A.}\ \bibnamefont
  {Curtiss}}\ and\ \bibinfo {author} {\bibfnamefont {J.~A.}\ \bibnamefont
  {Pople}},\ }\href {\doibase 10.1063/1.455788} {\bibfield  {journal} {\bibinfo
   {journal} {The Journal of Chemical Physics}\ }\textbf {\bibinfo {volume}
  {90}},\ \bibinfo {pages} {4314} (\bibinfo {year}
  {1989}{\natexlab{a}})}\BibitemShut {NoStop}%
\bibitem [{\citenamefont {Curtiss}\ and\ \citenamefont
  {Pople}(1989{\natexlab{b}})}]{Curtiss1989b}%
  \BibitemOpen
  \bibfield  {author} {\bibinfo {author} {\bibfnamefont {L.~A.}\ \bibnamefont
  {Curtiss}}\ and\ \bibinfo {author} {\bibfnamefont {J.~A.}\ \bibnamefont
  {Pople}},\ }\href {\doibase 10.1063/1.457605} {\bibfield  {journal} {\bibinfo
   {journal} {The Journal of Chemical Physics}\ }\textbf {\bibinfo {volume}
  {91}},\ \bibinfo {pages} {5118} (\bibinfo {year}
  {1989}{\natexlab{b}})}\BibitemShut {NoStop}%
\bibitem [{\citenamefont {Demachy}\ and\ \citenamefont
  {Volatron}(1994)}]{Demachy1994}%
  \BibitemOpen
  \bibfield  {author} {\bibinfo {author} {\bibfnamefont {I.}~\bibnamefont
  {Demachy}}\ and\ \bibinfo {author} {\bibfnamefont {F.}~\bibnamefont
  {Volatron}},\ }\href {\doibase 10.1021/j100093a010} {\bibfield  {journal}
  {\bibinfo  {journal} {The Journal of Physical Chemistry}\ }\textbf {\bibinfo
  {volume} {98}},\ \bibinfo {pages} {10728} (\bibinfo {year}
  {1994})}\BibitemShut {NoStop}%
\bibitem [{\citenamefont {Alkorta}\ \emph {et~al.}(2011)\citenamefont
  {Alkorta}, \citenamefont {Soteras}, \citenamefont {Elguero},\ and\
  \citenamefont {Bene}}]{Alkorta2011}%
  \BibitemOpen
  \bibfield  {author} {\bibinfo {author} {\bibfnamefont {I.}~\bibnamefont
  {Alkorta}}, \bibinfo {author} {\bibfnamefont {I.}~\bibnamefont {Soteras}},
  \bibinfo {author} {\bibfnamefont {J.}~\bibnamefont {Elguero}}, \ and\
  \bibinfo {author} {\bibfnamefont {J.~E.~D.}\ \bibnamefont {Bene}},\ }\href
  {\doibase 10.1039/c1cp20560a} {\bibfield  {journal} {\bibinfo  {journal}
  {Physical Chemistry Chemical Physics}\ }\textbf {\bibinfo {volume} {13}},\
  \bibinfo {pages} {14026} (\bibinfo {year} {2011})}\BibitemShut {NoStop}%
\bibitem [{\citenamefont {Power}(2010)}]{Power2010}%
  \BibitemOpen
  \bibfield  {author} {\bibinfo {author} {\bibfnamefont {P.~P.}\ \bibnamefont
  {Power}},\ }\href {\doibase 10.1038/nature08634} {\bibfield  {journal}
  {\bibinfo  {journal} {Nature}\ }\textbf {\bibinfo {volume} {463}},\ \bibinfo
  {pages} {171} (\bibinfo {year} {2010})}\BibitemShut {NoStop}%
\bibitem [{\citenamefont {Power}(2011)}]{Power2011}%
  \BibitemOpen
  \bibfield  {author} {\bibinfo {author} {\bibfnamefont {P.~P.}\ \bibnamefont
  {Power}},\ }\href {\doibase 10.1002/tcr.201100016} {\bibfield  {journal}
  {\bibinfo  {journal} {The Chemical Record}\ }\textbf {\bibinfo {volume}
  {12}},\ \bibinfo {pages} {238} (\bibinfo {year} {2011})}\BibitemShut
  {NoStop}%
\bibitem [{\citenamefont {Giffin}\ and\ \citenamefont
  {Masuda}(2011)}]{Giffin2011}%
  \BibitemOpen
  \bibfield  {author} {\bibinfo {author} {\bibfnamefont {N.~A.}\ \bibnamefont
  {Giffin}}\ and\ \bibinfo {author} {\bibfnamefont {J.~D.}\ \bibnamefont
  {Masuda}},\ }\href {\doibase 10.1016/j.ccr.2010.12.016} {\bibfield  {journal}
  {\bibinfo  {journal} {Coordination Chemistry Reviews}\ }\textbf {\bibinfo
  {volume} {255}},\ \bibinfo {pages} {1342} (\bibinfo {year}
  {2011})}\BibitemShut {NoStop}%
\bibitem [{\citenamefont {Niemeyer}\ and\ \citenamefont
  {Power}(1997)}]{Niemeyer1997}%
  \BibitemOpen
  \bibfield  {author} {\bibinfo {author} {\bibfnamefont {M.}~\bibnamefont
  {Niemeyer}}\ and\ \bibinfo {author} {\bibfnamefont {P.~P.}\ \bibnamefont
  {Power}},\ }\href {\doibase 10.1021/ic970319t} {\bibfield  {journal}
  {\bibinfo  {journal} {Inorganic Chemistry}\ }\textbf {\bibinfo {volume}
  {36}},\ \bibinfo {pages} {4688} (\bibinfo {year} {1997})}\BibitemShut
  {NoStop}%
\bibitem [{\citenamefont {Naglav}\ \emph {et~al.}(2015)\citenamefont {Naglav},
  \citenamefont {Neumann}, \citenamefont {Bl\"{a}ser}, \citenamefont
  {W\"{o}lper}, \citenamefont {Haack}, \citenamefont {Jansen},\ and\
  \citenamefont {Schulz}}]{Naglav2015}%
  \BibitemOpen
  \bibfield  {author} {\bibinfo {author} {\bibfnamefont {D.}~\bibnamefont
  {Naglav}}, \bibinfo {author} {\bibfnamefont {A.}~\bibnamefont {Neumann}},
  \bibinfo {author} {\bibfnamefont {D.}~\bibnamefont {Bl\"{a}ser}}, \bibinfo
  {author} {\bibfnamefont {C.}~\bibnamefont {W\"{o}lper}}, \bibinfo {author}
  {\bibfnamefont {R.}~\bibnamefont {Haack}}, \bibinfo {author} {\bibfnamefont
  {G.}~\bibnamefont {Jansen}}, \ and\ \bibinfo {author} {\bibfnamefont
  {S.}~\bibnamefont {Schulz}},\ }\href {\doibase 10.1039/c4cc09732g} {\bibfield
   {journal} {\bibinfo  {journal} {Chemical Communications}\ }\textbf {\bibinfo
  {volume} {51}},\ \bibinfo {pages} {3889} (\bibinfo {year}
  {2015})}\BibitemShut {NoStop}%
\bibitem [{\citenamefont {Arnold}\ \emph {et~al.}(2015)\citenamefont {Arnold},
  \citenamefont {Braunschweig}, \citenamefont {Ewing}, \citenamefont {Kramer},
  \citenamefont {Mies},\ and\ \citenamefont {Schuster}}]{Arnold2015}%
  \BibitemOpen
  \bibfield  {author} {\bibinfo {author} {\bibfnamefont {T.}~\bibnamefont
  {Arnold}}, \bibinfo {author} {\bibfnamefont {H.}~\bibnamefont
  {Braunschweig}}, \bibinfo {author} {\bibfnamefont {W.~C.}\ \bibnamefont
  {Ewing}}, \bibinfo {author} {\bibfnamefont {T.}~\bibnamefont {Kramer}},
  \bibinfo {author} {\bibfnamefont {J.}~\bibnamefont {Mies}}, \ and\ \bibinfo
  {author} {\bibfnamefont {J.~K.}\ \bibnamefont {Schuster}},\ }\href {\doibase
  10.1039/c4cc08519a} {\bibfield  {journal} {\bibinfo  {journal} {Chemical
  Communications}\ }\textbf {\bibinfo {volume} {51}},\ \bibinfo {pages} {737}
  (\bibinfo {year} {2015})}\BibitemShut {NoStop}%
\bibitem [{\citenamefont {Lerner}\ \emph {et~al.}(2003)\citenamefont {Lerner},
  \citenamefont {Scholz}, \citenamefont {Bolte}, \citenamefont {Wiberg},
  \citenamefont {N\"{o}th},\ and\ \citenamefont {Krossing}}]{Lerner2003}%
  \BibitemOpen
  \bibfield  {author} {\bibinfo {author} {\bibfnamefont {H.-W.}\ \bibnamefont
  {Lerner}}, \bibinfo {author} {\bibfnamefont {S.}~\bibnamefont {Scholz}},
  \bibinfo {author} {\bibfnamefont {M.}~\bibnamefont {Bolte}}, \bibinfo
  {author} {\bibfnamefont {N.}~\bibnamefont {Wiberg}}, \bibinfo {author}
  {\bibfnamefont {H.}~\bibnamefont {N\"{o}th}}, \ and\ \bibinfo {author}
  {\bibfnamefont {I.}~\bibnamefont {Krossing}},\ }\href {\doibase
  10.1002/ejic.200390092} {\bibfield  {journal} {\bibinfo  {journal} {European
  Journal of Inorganic Chemistry}\ }\textbf {\bibinfo {volume} {2003}},\
  \bibinfo {pages} {666} (\bibinfo {year} {2003})}\BibitemShut {NoStop}%
\bibitem [{\citenamefont {Mondal}\ \emph {et~al.}(2013)\citenamefont {Mondal},
  \citenamefont {Roesky}, \citenamefont {Schwarzer}, \citenamefont {Frenking},
  \citenamefont {Niep\"{o}tter}, \citenamefont {Wolf}, \citenamefont
  {Herbst-Irmer},\ and\ \citenamefont {Stalke}}]{Mondal2013}%
  \BibitemOpen
  \bibfield  {author} {\bibinfo {author} {\bibfnamefont {K.~C.}\ \bibnamefont
  {Mondal}}, \bibinfo {author} {\bibfnamefont {H.~W.}\ \bibnamefont {Roesky}},
  \bibinfo {author} {\bibfnamefont {M.~C.}\ \bibnamefont {Schwarzer}}, \bibinfo
  {author} {\bibfnamefont {G.}~\bibnamefont {Frenking}}, \bibinfo {author}
  {\bibfnamefont {B.}~\bibnamefont {Niep\"{o}tter}}, \bibinfo {author}
  {\bibfnamefont {H.}~\bibnamefont {Wolf}}, \bibinfo {author} {\bibfnamefont
  {R.}~\bibnamefont {Herbst-Irmer}}, \ and\ \bibinfo {author} {\bibfnamefont
  {D.}~\bibnamefont {Stalke}},\ }\href {\doibase 10.1002/anie.201208307}
  {\bibfield  {journal} {\bibinfo  {journal} {Angewandte Chemie International
  Edition}\ }\textbf {\bibinfo {volume} {52}},\ \bibinfo {pages} {2963}
  (\bibinfo {year} {2013})}\BibitemShut {NoStop}%
\bibitem [{\citenamefont {Li}\ \emph {et~al.}(2013)\citenamefont {Li},
  \citenamefont {Mondal}, \citenamefont {Roesky}, \citenamefont {Zhu},
  \citenamefont {Stollberg}, \citenamefont {Herbst-Irmer}, \citenamefont
  {Stalke},\ and\ \citenamefont {Andrada}}]{Li2013}%
  \BibitemOpen
  \bibfield  {author} {\bibinfo {author} {\bibfnamefont {Y.}~\bibnamefont
  {Li}}, \bibinfo {author} {\bibfnamefont {K.~C.}\ \bibnamefont {Mondal}},
  \bibinfo {author} {\bibfnamefont {H.~W.}\ \bibnamefont {Roesky}}, \bibinfo
  {author} {\bibfnamefont {H.}~\bibnamefont {Zhu}}, \bibinfo {author}
  {\bibfnamefont {P.}~\bibnamefont {Stollberg}}, \bibinfo {author}
  {\bibfnamefont {R.}~\bibnamefont {Herbst-Irmer}}, \bibinfo {author}
  {\bibfnamefont {D.}~\bibnamefont {Stalke}}, \ and\ \bibinfo {author}
  {\bibfnamefont {D.~M.}\ \bibnamefont {Andrada}},\ }\href {\doibase
  10.1021/ja406112u} {\bibfield  {journal} {\bibinfo  {journal} {Journal of the
  American Chemical Society}\ }\textbf {\bibinfo {volume} {135}},\ \bibinfo
  {pages} {12422} (\bibinfo {year} {2013})}\BibitemShut {NoStop}%
\bibitem [{\citenamefont {{\relax Sz}alay}(2015)}]{Szalay-2015b}%
  \BibitemOpen
  \bibfield  {author} {\bibinfo {author} {\bibfnamefont {{\relax
  Sz}.}~\bibnamefont {{\relax Sz}alay}},\ }\href {\doibase
  10.1103/PhysRevA.92.042329} {\bibfield  {journal} {\bibinfo  {journal} {Phys.
  Rev. A}\ }\textbf {\bibinfo {volume} {92}},\ \bibinfo {pages} {042329}
  (\bibinfo {year} {2015})}\BibitemShut {NoStop}%
\bibitem [{\citenamefont {Ohya}\ and\ \citenamefont {Petz}(1993)}]{Ohya-1993}%
  \BibitemOpen
  \bibfield  {author} {\bibinfo {author} {\bibfnamefont {M.}~\bibnamefont
  {Ohya}}\ and\ \bibinfo {author} {\bibfnamefont {D.}~\bibnamefont {Petz}},\
  }\href@noop {} {\emph {\bibinfo {title} {Quantum Entropy and Its Use}}},\
  \bibinfo {edition} {1st}\ ed.\ (\bibinfo  {publisher} {Springer Verlag},\
  \bibinfo {year} {1993})\BibitemShut {NoStop}%
\bibitem [{\citenamefont {Araki}\ and\ \citenamefont
  {Moriya}(2003)}]{Araki-2003b}%
  \BibitemOpen
  \bibfield  {author} {\bibinfo {author} {\bibfnamefont {H.}~\bibnamefont
  {Araki}}\ and\ \bibinfo {author} {\bibfnamefont {H.}~\bibnamefont {Moriya}},\
  }\href {\doibase 10.1142/S0129055X03001606} {\bibfield  {journal} {\bibinfo
  {journal} {Reviews in Mathematical Physics}\ }\textbf {\bibinfo {volume}
  {15}},\ \bibinfo {pages} {93} (\bibinfo {year} {2003})}\BibitemShut {NoStop}%
\bibitem [{\citenamefont {Wilde}(2013)}]{Wilde-2013}%
  \BibitemOpen
  \bibfield  {author} {\bibinfo {author} {\bibfnamefont {M.~M.}\ \bibnamefont
  {Wilde}},\ }\href@noop {} {\emph {\bibinfo {title} {Quantum Information
  Theory}}}\ (\bibinfo  {publisher} {Cambridge University Press},\ \bibinfo
  {year} {2013})\BibitemShut {NoStop}%
\bibitem [{\citenamefont {Horodecki}\ \emph {et~al.}(2009)\citenamefont
  {Horodecki}, \citenamefont {Horodecki}, \citenamefont {Horodecki},\ and\
  \citenamefont {Horodecki}}]{Horodecki-2009}%
  \BibitemOpen
  \bibfield  {author} {\bibinfo {author} {\bibfnamefont {R.}~\bibnamefont
  {Horodecki}}, \bibinfo {author} {\bibfnamefont {P.}~\bibnamefont
  {Horodecki}}, \bibinfo {author} {\bibfnamefont {M.}~\bibnamefont
  {Horodecki}}, \ and\ \bibinfo {author} {\bibfnamefont {K.}~\bibnamefont
  {Horodecki}},\ }\href {\doibase 10.1103/RevModPhys.81.865} {\bibfield
  {journal} {\bibinfo  {journal} {Rev. Mod. Phys.}\ }\textbf {\bibinfo {volume}
  {81}},\ \bibinfo {pages} {865} (\bibinfo {year} {2009})}\BibitemShut
  {NoStop}%
\bibitem [{\citenamefont {Schr{\"o}dinger}(1936)}]{Schrodinger-1936}%
  \BibitemOpen
  \bibfield  {author} {\bibinfo {author} {\bibfnamefont {E.}~\bibnamefont
  {Schr{\"o}dinger}},\ }\href {\doibase 10.1017/S0305004100019137} {\bibfield
  {journal} {\bibinfo  {journal} {Math. Proc. Camb. Phil. Soc.}\ }\textbf
  {\bibinfo {volume} {32}},\ \bibinfo {pages} {446} (\bibinfo {year}
  {1936})}\BibitemShut {NoStop}%
\bibitem [{\citenamefont {Hughston}\ \emph {et~al.}(1993)\citenamefont
  {Hughston}, \citenamefont {Jozsa},\ and\ \citenamefont
  {Wootters}}]{Hughston-1993}%
  \BibitemOpen
  \bibfield  {author} {\bibinfo {author} {\bibfnamefont {L.~P.}\ \bibnamefont
  {Hughston}}, \bibinfo {author} {\bibfnamefont {R.}~\bibnamefont {Jozsa}}, \
  and\ \bibinfo {author} {\bibfnamefont {W.~K.}\ \bibnamefont {Wootters}},\
  }\href {\doibase 10.1016/0375-9601(93)90880-9} {\bibfield  {journal}
  {\bibinfo  {journal} {Phys. Lett. A}\ }\textbf {\bibinfo {volume} {183}},\
  \bibinfo {pages} {14 } (\bibinfo {year} {1993})}\BibitemShut {NoStop}%
\bibitem [{\citenamefont {Davey}\ and\ \citenamefont
  {Priestley}(2002)}]{Davey-2002}%
  \BibitemOpen
  \bibfield  {author} {\bibinfo {author} {\bibfnamefont {B.~A.}\ \bibnamefont
  {Davey}}\ and\ \bibinfo {author} {\bibfnamefont {H.~A.}\ \bibnamefont
  {Priestley}},\ }\href@noop {} {\emph {\bibinfo {title} {Introduction to
  Lattices and Order}}},\ \bibinfo {edition} {2nd}\ ed.\ (\bibinfo  {publisher}
  {Cambridge University Press},\ \bibinfo {year} {2002})\BibitemShut {NoStop}%
\bibitem [{\citenamefont {{\relax Sz}alay}\ and\ \citenamefont
  {K\"ok\'enyesi}(2012)}]{Szalay-2012}%
  \BibitemOpen
  \bibfield  {author} {\bibinfo {author} {\bibfnamefont {{\relax
  Sz}.}~\bibnamefont {{\relax Sz}alay}}\ and\ \bibinfo {author} {\bibfnamefont
  {Z.}~\bibnamefont {K\"ok\'enyesi}},\ }\href {\doibase
  10.1103/PhysRevA.86.032341} {\bibfield  {journal} {\bibinfo  {journal} {Phys.
  Rev. A}\ }\textbf {\bibinfo {volume} {86}},\ \bibinfo {pages} {032341}
  (\bibinfo {year} {2012})}\BibitemShut {NoStop}%
\bibitem [{\citenamefont {Szalay}(2018)}]{Szalay2018cc}%
  \BibitemOpen
  \bibfield  {author} {\bibinfo {author} {\bibfnamefont {S.}~\bibnamefont
  {Szalay}},\ }\href {\doibase 10.1088/1751-8121/aae971} {\bibfield  {journal}
  {\bibinfo  {journal} {Journal of Physics A: Mathematical and Theoretical}\
  }\textbf {\bibinfo {volume} {51}},\ \bibinfo {pages} {485302} (\bibinfo
  {year} {2018})}\BibitemShut {NoStop}%
\bibitem [{\citenamefont {Adesso}\ \emph {et~al.}(2016)\citenamefont {Adesso},
  \citenamefont {Bromley},\ and\ \citenamefont {Cianciaruso}}]{Adesso-2016}%
  \BibitemOpen
  \bibfield  {author} {\bibinfo {author} {\bibfnamefont {G.}~\bibnamefont
  {Adesso}}, \bibinfo {author} {\bibfnamefont {T.~R.}\ \bibnamefont {Bromley}},
  \ and\ \bibinfo {author} {\bibfnamefont {M.}~\bibnamefont {Cianciaruso}},\
  }\href {http://stacks.iop.org/1751-8121/49/i=47/a=473001} {\bibfield
  {journal} {\bibinfo  {journal} {Journal of Physics A: Mathematical and
  Theoretical}\ }\textbf {\bibinfo {volume} {49}},\ \bibinfo {pages} {473001}
  (\bibinfo {year} {2016})}\BibitemShut {NoStop}%
\bibitem [{\citenamefont {Legeza}\ and\ \citenamefont
  {S{\'o}lyom}(2006)}]{Legeza-2006a}%
  \BibitemOpen
  \bibfield  {author} {\bibinfo {author} {\bibfnamefont {{\"O}.}~\bibnamefont
  {Legeza}}\ and\ \bibinfo {author} {\bibfnamefont {J.}~\bibnamefont
  {S{\'o}lyom}},\ }\href {\doibase 10.1103/PhysRevLett.96.116401} {\bibfield
  {journal} {\bibinfo  {journal} {Phys. Rev. Lett.}\ }\textbf {\bibinfo
  {volume} {96}},\ \bibinfo {pages} {116401} (\bibinfo {year}
  {2006})}\BibitemShut {NoStop}%
\bibitem [{\citenamefont {Rissler}\ \emph
  {et~al.}(2006{\natexlab{b}})\citenamefont {Rissler}, \citenamefont {Noack},\
  and\ \citenamefont {White}}]{Rissler-2006}%
  \BibitemOpen
  \bibfield  {author} {\bibinfo {author} {\bibfnamefont {J.}~\bibnamefont
  {Rissler}}, \bibinfo {author} {\bibfnamefont {R.~M.}\ \bibnamefont {Noack}},
  \ and\ \bibinfo {author} {\bibfnamefont {S.~R.}\ \bibnamefont {White}},\
  }\href {\doibase 10.1016/j.chemphys.2005.10.018} {\bibfield  {journal}
  {\bibinfo  {journal} {Chemical Physics}\ }\textbf {\bibinfo {volume} {323}},\
  \bibinfo {pages} {519 } (\bibinfo {year} {2006}{\natexlab{b}})}\BibitemShut
  {NoStop}%
\bibitem [{\citenamefont {Mottet}\ \emph {et~al.}(2014)\citenamefont {Mottet},
  \citenamefont {Tecmer}, \citenamefont {Boguslawski}, \citenamefont {Legeza},\
  and\ \citenamefont {Reiher}}]{Mottet-2014}%
  \BibitemOpen
  \bibfield  {author} {\bibinfo {author} {\bibfnamefont {M.}~\bibnamefont
  {Mottet}}, \bibinfo {author} {\bibfnamefont {P.}~\bibnamefont {Tecmer}},
  \bibinfo {author} {\bibfnamefont {K.}~\bibnamefont {Boguslawski}}, \bibinfo
  {author} {\bibfnamefont {{\"O}.}~\bibnamefont {Legeza}}, \ and\ \bibinfo
  {author} {\bibfnamefont {M.}~\bibnamefont {Reiher}},\ }\href {\doibase
  10.1039/C4CP00277F} {\bibfield  {journal} {\bibinfo  {journal} {Phys. Chem.
  Chem. Phys.}\ }\textbf {\bibinfo {volume} {16}},\ \bibinfo {pages} {8872}
  (\bibinfo {year} {2014})}\BibitemShut {NoStop}%
\bibitem [{\citenamefont {Murg}\ \emph
  {et~al.}(2015{\natexlab{b}})\citenamefont {Murg}, \citenamefont {Verstraete},
  \citenamefont {Schneider}, \citenamefont {Nagy},\ and\ \citenamefont
  {Legeza}}]{Murg-2015}%
  \BibitemOpen
  \bibfield  {author} {\bibinfo {author} {\bibfnamefont {V.}~\bibnamefont
  {Murg}}, \bibinfo {author} {\bibfnamefont {F.}~\bibnamefont {Verstraete}},
  \bibinfo {author} {\bibfnamefont {R.}~\bibnamefont {Schneider}}, \bibinfo
  {author} {\bibfnamefont {P.~R.}\ \bibnamefont {Nagy}}, \ and\ \bibinfo
  {author} {\bibfnamefont {{\"O}.}~\bibnamefont {Legeza}},\ }\href {\doibase
  10.1021/ct501187j} {\bibfield  {journal} {\bibinfo  {journal} {Journal of
  Chemical Theory and Computation}\ }\textbf {\bibinfo {volume} {11}},\
  \bibinfo {pages} {1027} (\bibinfo {year} {2015}{\natexlab{b}})}\BibitemShut
  {NoStop}%
\bibitem [{\citenamefont {Barcza}\ \emph {et~al.}(2015)\citenamefont {Barcza},
  \citenamefont {Noack}, \citenamefont {S{\'o}lyom},\ and\ \citenamefont
  {Legeza}}]{Barcza-2015}%
  \BibitemOpen
  \bibfield  {author} {\bibinfo {author} {\bibfnamefont {G.}~\bibnamefont
  {Barcza}}, \bibinfo {author} {\bibfnamefont {R.~M.}\ \bibnamefont {Noack}},
  \bibinfo {author} {\bibfnamefont {J.}~\bibnamefont {S{\'o}lyom}}, \ and\
  \bibinfo {author} {\bibfnamefont {{\"O}.}~\bibnamefont {Legeza}},\ }\href
  {\doibase 10.1103/PhysRevB.92.125140} {\bibfield  {journal} {\bibinfo
  {journal} {Phys. Rev. B}\ }\textbf {\bibinfo {volume} {92}},\ \bibinfo
  {pages} {125140} (\bibinfo {year} {2015})}\BibitemShut {NoStop}%
\bibitem [{\citenamefont {Lindblad}(1973)}]{Lindblad-1973}%
  \BibitemOpen
  \bibfield  {author} {\bibinfo {author} {\bibfnamefont {G.}~\bibnamefont
  {Lindblad}},\ }\href@noop {} {\bibfield  {journal} {\bibinfo  {journal}
  {Communications in Mathematical Physics}\ }\textbf {\bibinfo {volume} {33}},\
  \bibinfo {pages} {305} (\bibinfo {year} {1973})}\BibitemShut {NoStop}%
\bibitem [{\citenamefont {Horodecki}(1994)}]{Horodecki-1994}%
  \BibitemOpen
  \bibfield  {author} {\bibinfo {author} {\bibfnamefont {R.}~\bibnamefont
  {Horodecki}},\ }\href {\doibase 10.1016/0375-9601(94)90052-3} {\bibfield
  {journal} {\bibinfo  {journal} {Physics Letters A}\ }\textbf {\bibinfo
  {volume} {187}},\ \bibinfo {pages} {145 } (\bibinfo {year}
  {1994})}\BibitemShut {NoStop}%
\bibitem [{\citenamefont {Legeza}\ \emph {et~al.}(2006)\citenamefont {Legeza},
  \citenamefont {Gebhard},\ and\ \citenamefont {Rissler}}]{Legeza-2006b}%
  \BibitemOpen
  \bibfield  {author} {\bibinfo {author} {\bibfnamefont {{\"O}.}~\bibnamefont
  {Legeza}}, \bibinfo {author} {\bibfnamefont {F.}~\bibnamefont {Gebhard}}, \
  and\ \bibinfo {author} {\bibfnamefont {J.}~\bibnamefont {Rissler}},\ }\href
  {\doibase 10.1103/PhysRevB.74.195112} {\bibfield  {journal} {\bibinfo
  {journal} {Phys. Rev. B}\ }\textbf {\bibinfo {volume} {74}},\ \bibinfo
  {pages} {195112} (\bibinfo {year} {2006})}\BibitemShut {NoStop}%
\bibitem [{\citenamefont {Herbut}(2004)}]{Herbut-2004}%
  \BibitemOpen
  \bibfield  {author} {\bibinfo {author} {\bibfnamefont {F.}~\bibnamefont
  {Herbut}},\ }\href {http://stacks.iop.org/0305-4470/37/i=10/a=016} {\bibfield
   {journal} {\bibinfo  {journal} {Journal of Physics A: Mathematical and
  General}\ }\textbf {\bibinfo {volume} {37}},\ \bibinfo {pages} {3535}
  (\bibinfo {year} {2004})}\BibitemShut {NoStop}%
\bibitem [{\citenamefont {Modi}\ \emph {et~al.}(2010)\citenamefont {Modi},
  \citenamefont {Paterek}, \citenamefont {Son}, \citenamefont {Vedral},\ and\
  \citenamefont {Williamson}}]{Modi-2010}%
  \BibitemOpen
  \bibfield  {author} {\bibinfo {author} {\bibfnamefont {K.}~\bibnamefont
  {Modi}}, \bibinfo {author} {\bibfnamefont {T.}~\bibnamefont {Paterek}},
  \bibinfo {author} {\bibfnamefont {W.}~\bibnamefont {Son}}, \bibinfo {author}
  {\bibfnamefont {V.}~\bibnamefont {Vedral}}, \ and\ \bibinfo {author}
  {\bibfnamefont {M.}~\bibnamefont {Williamson}},\ }\href {\doibase
  10.1103/PhysRevLett.104.080501} {\bibfield  {journal} {\bibinfo  {journal}
  {Phys. Rev. Lett.}\ }\textbf {\bibinfo {volume} {104}},\ \bibinfo {pages}
  {080501} (\bibinfo {year} {2010})}\BibitemShut {NoStop}%
\bibitem [{\citenamefont {Bennett}\ \emph {et~al.}(1996)\citenamefont
  {Bennett}, \citenamefont {Bernstein}, \citenamefont {Popescu},\ and\
  \citenamefont {Schumacher}}]{Bennett-1996}%
  \BibitemOpen
  \bibfield  {author} {\bibinfo {author} {\bibfnamefont {C.~H.}\ \bibnamefont
  {Bennett}}, \bibinfo {author} {\bibfnamefont {H.~J.}\ \bibnamefont
  {Bernstein}}, \bibinfo {author} {\bibfnamefont {S.}~\bibnamefont {Popescu}},
  \ and\ \bibinfo {author} {\bibfnamefont {B.}~\bibnamefont {Schumacher}},\
  }\href {\doibase 10.1103/PhysRevA.53.2046} {\bibfield  {journal} {\bibinfo
  {journal} {Phys. Rev. A}\ }\textbf {\bibinfo {volume} {53}},\ \bibinfo
  {pages} {2046} (\bibinfo {year} {1996})}\BibitemShut {NoStop}%
\bibitem [{\citenamefont {Nielsen}\ and\ \citenamefont
  {Chuang}(2000)}]{Nielsen-2000}%
  \BibitemOpen
  \bibfield  {author} {\bibinfo {author} {\bibfnamefont {M.~A.}\ \bibnamefont
  {Nielsen}}\ and\ \bibinfo {author} {\bibfnamefont {I.~L.}\ \bibnamefont
  {Chuang}},\ }\href@noop {} {\emph {\bibinfo {title} {Quantum Computation and
  Quantum Information}}},\ \bibinfo {edition} {1st}\ ed.\ (\bibinfo
  {publisher} {Cambridge University Press},\ \bibinfo {year}
  {2000})\BibitemShut {NoStop}%
\bibitem [{\citenamefont {Schmidt}(1907)}]{Schmidt-1907}%
  \BibitemOpen
  \bibfield  {author} {\bibinfo {author} {\bibfnamefont {E.}~\bibnamefont
  {Schmidt}},\ }\href {http://eudml.org/doc/158296} {\bibfield  {journal}
  {\bibinfo  {journal} {Math. Ann.}\ }\textbf {\bibinfo {volume} {63}},\
  \bibinfo {pages} {433} (\bibinfo {year} {1907})}\BibitemShut {NoStop}%
\bibitem [{\citenamefont {Pipek}\ and\ \citenamefont
  {Mezey}(1989)}]{Pipek-1989}%
  \BibitemOpen
  \bibfield  {author} {\bibinfo {author} {\bibfnamefont {J.}~\bibnamefont
  {Pipek}}\ and\ \bibinfo {author} {\bibfnamefont {P.~G.}\ \bibnamefont
  {Mezey}},\ }\href {\doibase 10.1063/1.456588} {\bibfield  {journal} {\bibinfo
   {journal} {The Journal of Chemical Physics}\ }\textbf {\bibinfo {volume}
  {90}},\ \bibinfo {pages} {4916} (\bibinfo {year} {1989})}\BibitemShut
  {NoStop}%
\bibitem [{\citenamefont {Werner}\ \emph {et~al.}(2010)\citenamefont {Werner},
  \citenamefont {Knowles}, \citenamefont {Knizia}, \citenamefont {Manby},\ and\
  \citenamefont {Sch\"utz}}]{MOLPRO}%
  \BibitemOpen
  \bibfield  {author} {\bibinfo {author} {\bibfnamefont {H.~J.}\ \bibnamefont
  {Werner}}, \bibinfo {author} {\bibfnamefont {P.~J.}\ \bibnamefont {Knowles}},
  \bibinfo {author} {\bibfnamefont {G.}~\bibnamefont {Knizia}}, \bibinfo
  {author} {\bibfnamefont {F.~R.}\ \bibnamefont {Manby}}, \ and\ \bibinfo
  {author} {\bibfnamefont {M.}~\bibnamefont {Sch\"utz}},\ }\href@noop {}
  {\enquote {\bibinfo {title} {Molpro, version 2010.1, a package of ab initio
  programs, \texttt{http://www.molpro.net}},}\ } (\bibinfo {year}
  {2010})\BibitemShut {NoStop}%
\bibitem [{\citenamefont {Legeza}\ \emph {et~al.}()\citenamefont {Legeza},
  \citenamefont {Veis},\ and\ \citenamefont {Mosoni}}]{budapest_qcdmrg}%
  \BibitemOpen
  \bibfield  {author} {\bibinfo {author} {\bibfnamefont {{\"O}.}~\bibnamefont
  {Legeza}}, \bibinfo {author} {\bibfnamefont {L.}~\bibnamefont {Veis}}, \ and\
  \bibinfo {author} {\bibfnamefont {T.}~\bibnamefont {Mosoni}},\ }\href@noop {}
  {\enquote {\bibinfo {title} {{QC-DMRG-Budapest, a program for quantum
  chemical DMRG calculations}},}\ }\BibitemShut {NoStop}%
\bibitem [{\citenamefont {Chalupsky}()}]{charmol}%
  \BibitemOpen
  \bibfield  {author} {\bibinfo {author} {\bibfnamefont {J.}~\bibnamefont
  {Chalupsky}},\ }\href@noop {} {\enquote {\bibinfo {title} {Charmol: program
  for molecular graphics},}\ }\bibinfo {howpublished}
  {\url{https://sourceforge.net/projects/charmol}},\ \bibinfo {note} {accessed:
  2018-09-09}\BibitemShut {NoStop}%
\bibitem [{\citenamefont {Osborne}\ and\ \citenamefont
  {Verstraete}(2006)}]{Osborne2006}%
  \BibitemOpen
  \bibfield  {author} {\bibinfo {author} {\bibfnamefont {T.~J.}\ \bibnamefont
  {Osborne}}\ and\ \bibinfo {author} {\bibfnamefont {F.}~\bibnamefont
  {Verstraete}},\ }\href {\doibase 10.1103/physrevlett.96.220503} {\bibfield
  {journal} {\bibinfo  {journal} {Physical Review Letters}\ }\textbf {\bibinfo
  {volume} {96}} (\bibinfo {year} {2006}),\
  10.1103/physrevlett.96.220503}\BibitemShut {NoStop}%
\bibitem [{\citenamefont {Coffman}\ \emph {et~al.}(2000)\citenamefont
  {Coffman}, \citenamefont {Kundu},\ and\ \citenamefont
  {Wootters}}]{Coffman2000}%
  \BibitemOpen
  \bibfield  {author} {\bibinfo {author} {\bibfnamefont {V.}~\bibnamefont
  {Coffman}}, \bibinfo {author} {\bibfnamefont {J.}~\bibnamefont {Kundu}}, \
  and\ \bibinfo {author} {\bibfnamefont {W.~K.}\ \bibnamefont {Wootters}},\
  }\href {\doibase 10.1103/physreva.61.052306} {\bibfield  {journal} {\bibinfo
  {journal} {Physical Review A}\ }\textbf {\bibinfo {volume} {61}} (\bibinfo
  {year} {2000}),\ 10.1103/physreva.61.052306}\BibitemShut {NoStop}%
\bibitem [{\citenamefont {Szilvasi}\ \emph {et~al.}(2015)\citenamefont
  {Szilvasi}, \citenamefont {Barcza},\ and\ \citenamefont
  {Legeza}}]{szilvasi_2015}%
  \BibitemOpen
  \bibfield  {author} {\bibinfo {author} {\bibfnamefont {T.}~\bibnamefont
  {Szilvasi}}, \bibinfo {author} {\bibfnamefont {G.}~\bibnamefont {Barcza}}, \
  and\ \bibinfo {author} {\bibfnamefont {O.}~\bibnamefont {Legeza}},\ }\href
  {http://arxiv.org/abs/1509.04241} {\bibfield  {journal} {\bibinfo  {journal}
  {arXiv [quant-ph]}\ ,\ \bibinfo {eid} {1509.04241}} (\bibinfo {year}
  {2015})}\BibitemShut {NoStop}%
\bibitem [{\citenamefont {Krapp}\ \emph {et~al.}(2007)\citenamefont {Krapp},
  \citenamefont {Pandey},\ and\ \citenamefont {Frenking}}]{Krapp2007}%
  \BibitemOpen
  \bibfield  {author} {\bibinfo {author} {\bibfnamefont {A.}~\bibnamefont
  {Krapp}}, \bibinfo {author} {\bibfnamefont {K.~K.}\ \bibnamefont {Pandey}}, \
  and\ \bibinfo {author} {\bibfnamefont {G.}~\bibnamefont {Frenking}},\ }\href
  {\doibase 10.1021/ja0691324} {\bibfield  {journal} {\bibinfo  {journal} {J.
  Am. Chem. Soc.}\ }\textbf {\bibinfo {volume} {129}},\ \bibinfo {pages} {7596}
  (\bibinfo {year} {2007})}\BibitemShut {NoStop}%
\end{thebibliography}%

\appendix
\section{Eigenvectors of the reduced density operators}\label{appendix}

\subsection{Be(CAC)$_{2}$}
The (reduced) density operator
of the $X_1$ orbitals
consists of the following eigenstates of the three highest eigenvalues (probabilities).\\
Probability $0.5798$:
\begin{align*}
  \hskip -0.2cm
  \ket{\psi_{X_1}^1} =&+0.0864 \ket{--\updownarrows}
  +0.3255 \ket{-\downarrow\;\uparrow}
  -0.3255 \ket{-\uparrow\;\downarrow} \\
  & +0.3324 \ket{-\updownarrows-}
  +0.4748 \ket{\downarrow-\uparrow}
  -0.3254 \ket{\downarrow\;\uparrow-} \\
  & -0.4748 \ket{\uparrow-\downarrow}
  +0.3254 \ket{\uparrow\;\downarrow-}
  +0.0863 \ket{\updownarrows--}
\end{align*}
Probability $0.1473$:
\begin{align}
  \hskip -0.2cm
  \ket{\psi_{X_1}^2} =&-0.2010 \ket{-\downarrow\;\updownarrows}
  -0.2630 \ket{-\updownarrows\;\downarrow}
  -0.3980 \ket{\downarrow-\updownarrows}\nonumber \\
  & -0.2781 \ket{\downarrow\;\downarrow\;\uparrow}
  +0.5562 \ket{\downarrow\;\uparrow\;\downarrow}
  -0.2630 \ket{\downarrow\;\updownarrows-}\nonumber \\
  & -0.2781 \ket{\uparrow\;\downarrow\;\downarrow}
  -0.3980 \ket{\updownarrows-\downarrow}
  +0.2010 \ket{\updownarrows\downarrow-}\nonumber
\end{align}
Probability $0.1473$:
\begin{align}
  \hskip -0.2cm
  \ket{\psi_{X_1}^3} =&+0.2010 \ket{-\uparrow\;\updownarrows}
  +0.2630 \ket{-\updownarrows\;\uparrow}
  -0.2781 \ket{\downarrow\;\uparrow\;\uparrow}\nonumber \\
  & +0.3980 \ket{\uparrow-\updownarrows}
  +0.5562 \ket{\uparrow\;\downarrow\;\uparrow}
  -0.2781 \ket{\uparrow\;\uparrow\;\downarrow}\nonumber \\
  & +0.2630 \ket{\uparrow\;\updownarrows-}
  +0.3980 \ket{\updownarrows-\uparrow}
  -0.2010 \ket{\updownarrows\;\uparrow-}\nonumber
\end{align}
all the other eigenvalues are less than $0.033$.

\subsection{[Be(CAC)$_{2}$]$^{2+}$}
As mentioned earlier in the text, the reduced density operator
for $X_1$ is highly mixed in this this case, with no dominant state.
Therefore we only list the highest eigenvalues to show this:
\begin{align}
\hskip -0.2cm
\;0.1016,\;
0.1007,\;
0.1007,\;
0.1007,\;
0.1002,\nonumber \\
0.1002,\;
0.0899,\;
0.0899,\;
0.0868,\;
0.0214.\nonumber
\end{align}

\subsection{Distribution of the two-orbital correlations}
\begin{figure*}[!ht]
	\hspace*{\fill}
  \subfloat[diborane(6)\label{Id6}]{%
    \includegraphics[width=5.33cm]{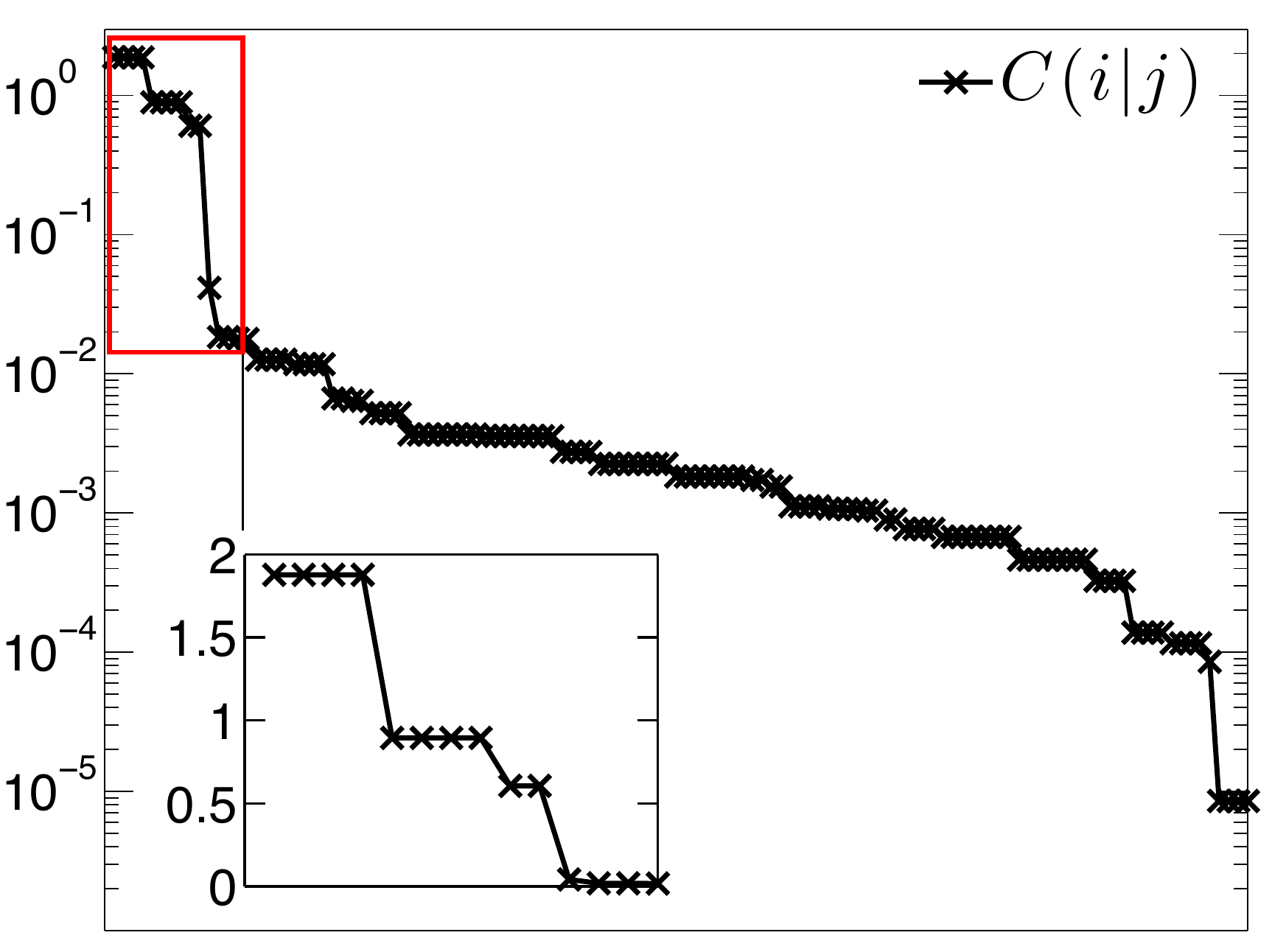}
  }
  \hfill
  \subfloat[diborane(4)\label{Id4}]{%
    \includegraphics[width=5.33cm]{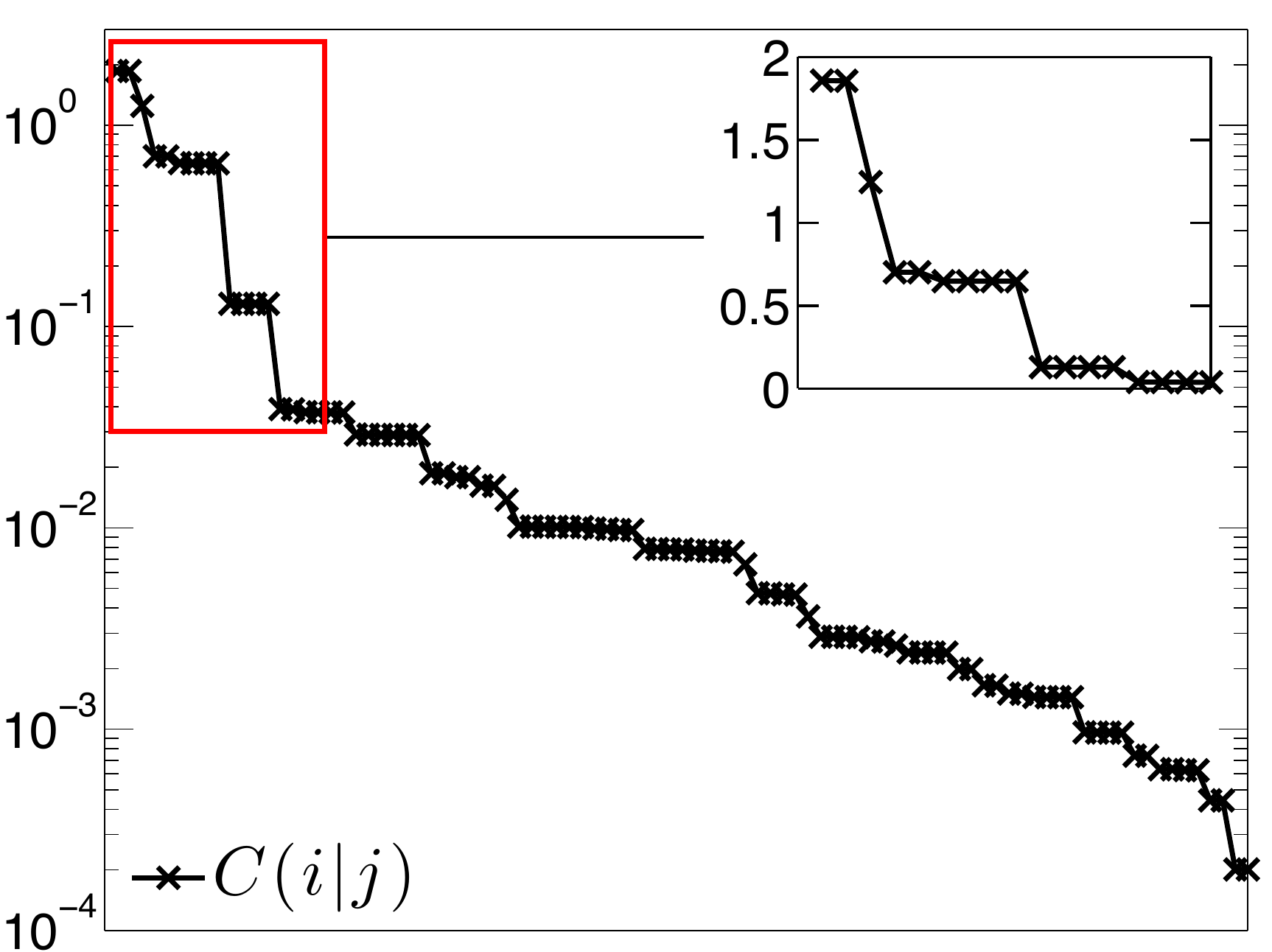}
  }
  \hspace*{\fill}
  \\
  \subfloat[Be(CAC)$_{2}$\label{Icbecneutral}]{%
    \includegraphics[width=5.33cm]{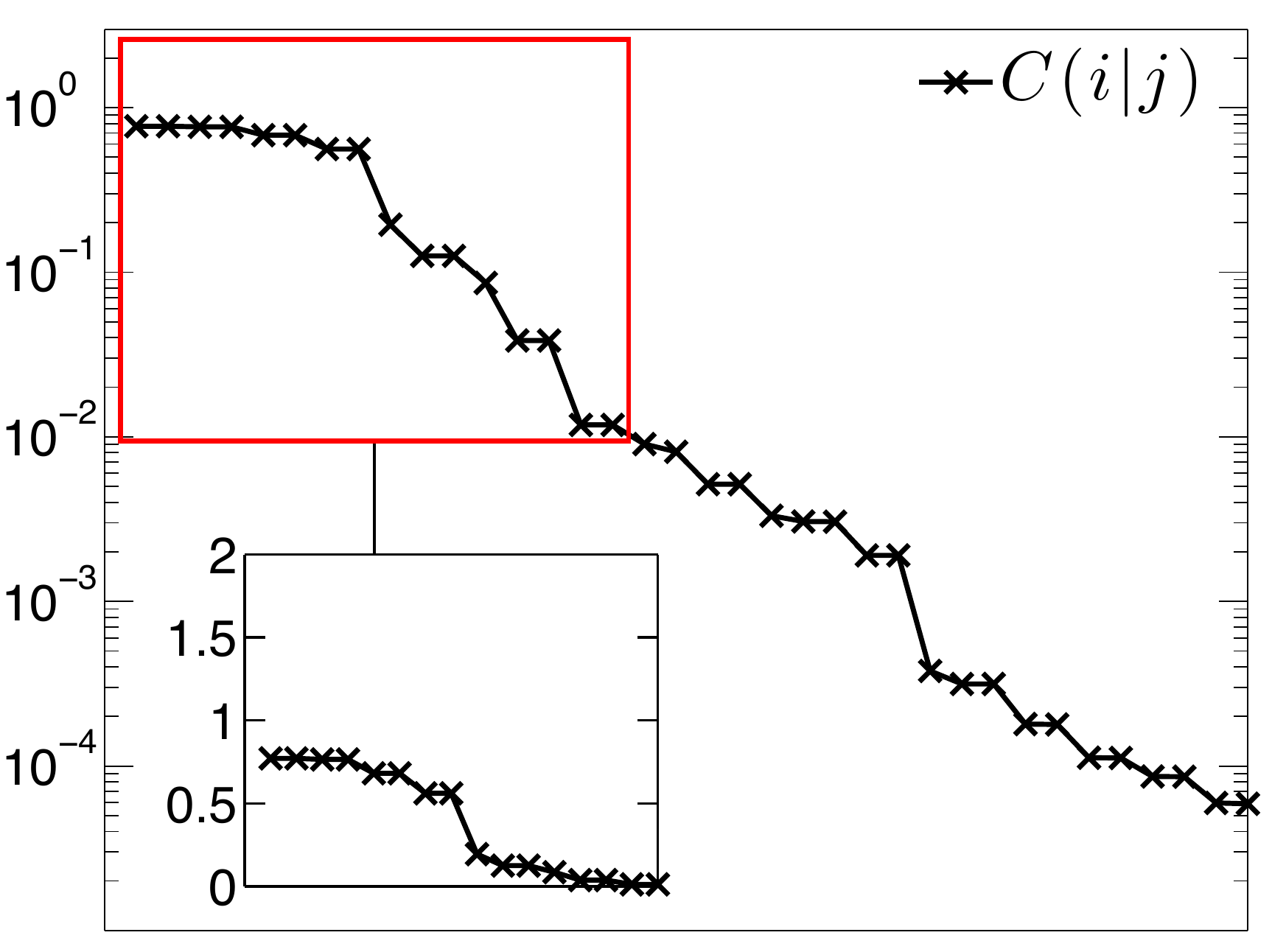}
  }
  \hfill
	\subfloat[ ${[ \text{Be(CAC)}_2 ]}^{2+}$ \label{Icbec2plus}]{%
    \includegraphics[width=5.33cm]{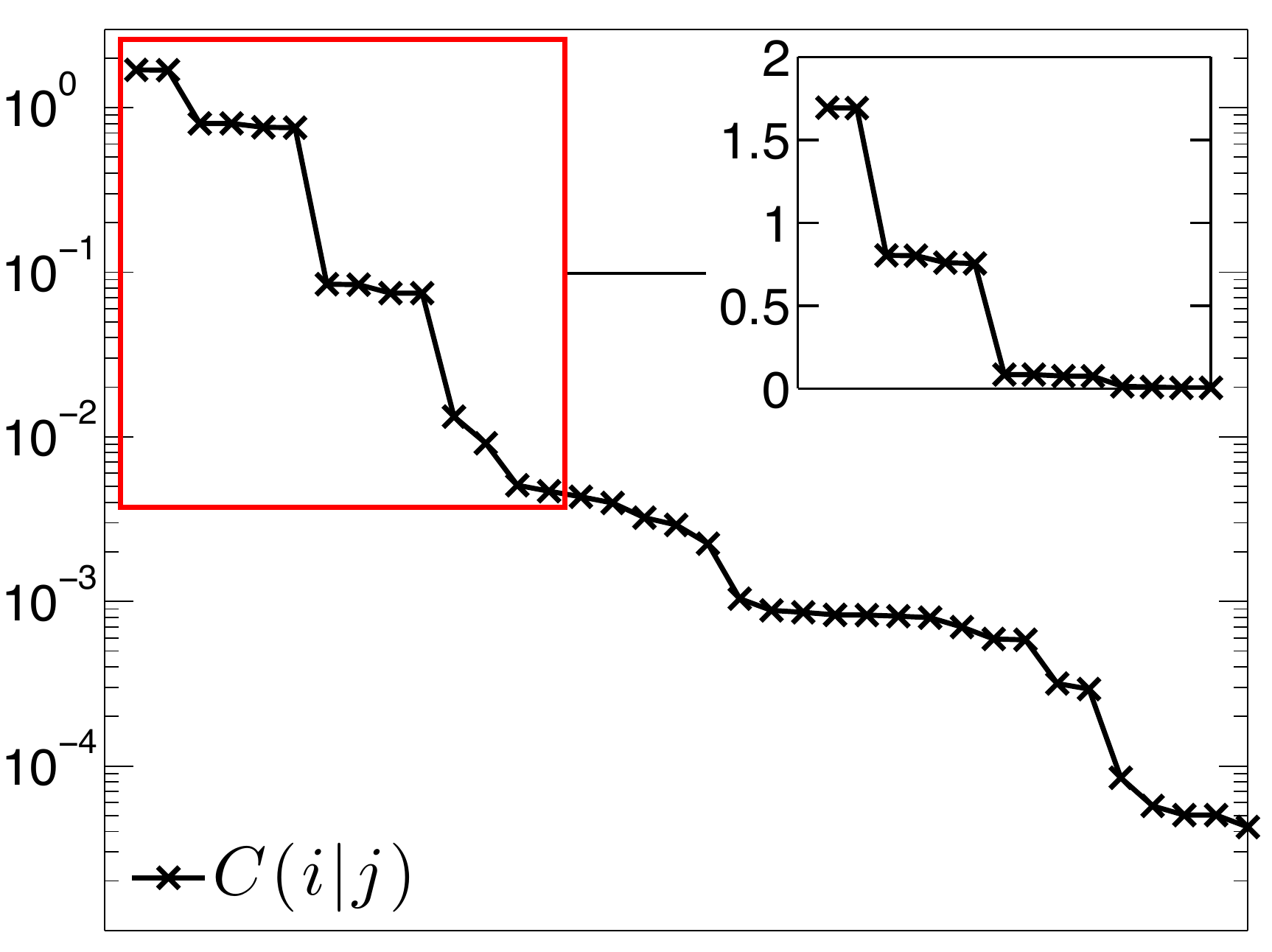}
  }
  \hfill
  \subfloat[Be(CAC)$_{2}$ with rotated $X_2$\label{Icbecrotated}]{%
    \includegraphics[width=5.33cm]{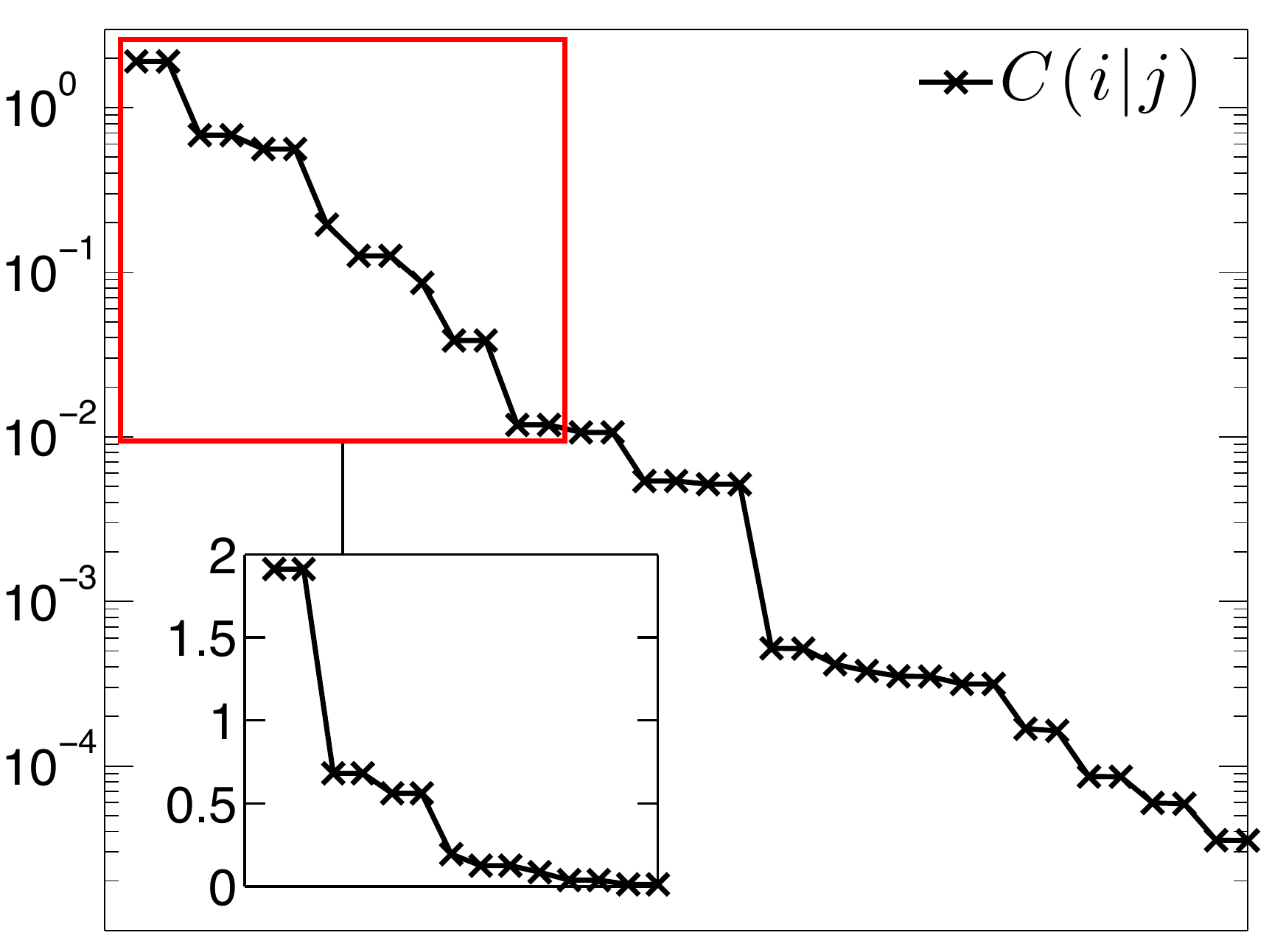}
  }
  \caption{The distributions of the two-orbital correlations for diborane and beryllium complexes.}\label{decay}
\end{figure*}

Figure \ref{decay} shows the distribution of the two-orbital correlations for diborane and beryllium complexes.

\end{document}